\documentstyle[preprint,tighten,12pt,aps]{revtex}
\newcommand{\bsigma} {{\mbox{\boldmath $\sigma$}}}

\newcommand{\bnabla} {{\mbox{\boldmath $\nabla$}}}
\newcommand{\bdel} {{\mbox{\boldmath $\nabla$}}}
\newcommand{\bPi}{{\mbox{\boldmath $\Pi$}}}
\def\A{{\bf A}}  
\def\B{{\bf B}}
\def\x{{\bf x}}
\def\y{{\bf y}}
\def\k{{\bf k}}
\def\s{{\bf s}}
\def\l{{\bf l}}
\def\q{{\bf q}}
\def\z{{\bf z}}
\def\D{{\bf D}}
\def\P{{\bf P}}
\def\p{{\bf p}}
\def\E{{\bf E}}
\def\dk{ { {d{\bf k} } \over {(2\pi)^3}  }}
\def\dq{ { {d{\bf q} } \over {(2\pi)^3}  }}
\def\ddp{ { {d{\bf p} } \over {(2\pi)^3}  }}

\newcommand{\gapproxeq}{\lower.7ex\hbox{$\;\stackrel{\textstyle>}{\sim}\;$}}
\newcommand{\lapproxeq}{\lower.7ex\hbox{$\;\stackrel{\textstyle<}{\sim}\;$}}

\begin{document}
\voffset = 0.3 true in
\input{psfig.sty}
\title{Coulomb Gauge QCD, Confinement, and the Constituent Representation}

\author{Adam P. Szczepaniak$^1$ and  Eric S. Swanson$^{2}$ }

\address{$^1$\
  Department of Physics and Nuclear Theory Center \\
  Indiana University, 
  Bloomington, Indiana   47405-4202\\
         $^2$
Department of Physics and Astronomy, University of Pittsburgh,
Pittsburgh PA 15260 and \\ 
Jefferson Lab, 12000 Jefferson Ave,
Newport News, VA 23606.         
        }

\maketitle

\begin{abstract}
Quark confinement and the genesis of the constituent quark
model are examined in nonperturbative QCD in Coulomb gauge.
We employ a self-consistent method to construct a quasiparticle
basis and to determine the quasiparticle interaction. The results 
agree remarkably well with lattice computations. They
also illustrate the mechanism by which confinement and 
constituent quarks emerge, provide support for the Gribov-Zwanziger
confinement scenario,
clarify several perplexing issues in the constituent quark model, and
permit the construction of an improved model of low energy QCD.


\end{abstract}



\section{Introduction}

Two of the key issues facing QCD at low energy are a quantitative description
of confinement and an understanding of the  origins of the constituent 
 quark model.
In this paper we demonstrate how both issues may be resolved through a nonperturbative
analysis of QCD in Coulomb gauge. This demonstration makes the physical origin
of both effects clear,
resolves several longstanding inconsistencies in the constituent quark model, 
significantly extends the quark model, and 
establishes a -- perhaps surprising -- relationship between confinement and
the constituent quark model.

Although lattice gauge computations are capable of answering many questions
in strong QCD, it is clear that the development of
reliable analytical continuum tools are a necessity for advancing the 
field\cite{Capstick:2000dk}. Continuum methods allow one to
understand how QCD works from first principles, permit the
development of intuition for phenomenological model building, and address
computationally challenging phenomena such as QCD at finite density, 
 extrapolation to low quark  
masses, or the treatment of large hadronic systems. A variety of such
 continuum  
 tools exist: chiral perturbation
 theory, effective heavy quark and low energy hadronic field theories, 
 4-dimensional Dyson-Schwinger 
 methods, fixed gauge Hamiltonian QCD approaches, and QCD sum rule methods.
In this paper we focus on Hamiltonian QCD in Coulomb gauge. 

Much progress has been made in understanding Coulomb gauge QCD 
since the seminal work of Schwinger\cite{Schwinger}, Khriplovich\cite{Khrip}, 
and Christ and Lee\cite{Christ:1980ku}.
In particular, the problem of the Gribov
ambiguity\cite{gribov} has been studied and a resolution has been 
suggested\cite{Zwanziger:1998ez}.
The ambiguity  arises because of residual gauge
freedom  after the canonical Coulomb gauge fixing condition, 
$\nabla\cdot\A = 0$ is imposed in a nonabelian theory. 
In Ref.~\cite{gribov} Gribov noted that the multiple-gauge copy 
ambiguity may be resolved by insisting that the Faddeev-Popov 
operator (to be defined later) is positive. With the aid of a simple model,
he then showed that this constraint implies the existence of a 
novel form for the gluon
propagator and an 
enhancement in the Faddeev-Popov propagator at low momenta. Furthermore, these 
imply that an 
enhancement exists in the instantaneous Coulomb  potential, thereby providing a 
plausible mechanism for confinement.
In a series of recent
papers~\cite{Zwanziger:1998ez,Cucchieri:1997ja}, Zwanziger has 
brought the Gribov Coulomb gauge
confinement scenario onto firm theoretical ground and has demonstrated
that a complete definition of the Coulomb gauge may be achieved by restricting the gauge
fields to the `fundamental modular region' -- defined as the set of gauge fields which form
the absolute minima of a suitable functional. Furthermore, the constraint to the fundamental modular region
may be 
imposed by introducing a horizon term
through a Lagrange multiplier in the Hamiltonian.

A key feature of Coulomb gauge is that the elimination of 
nondynamical degrees of
freedom creates an instantaneous interaction. The QED analogue of this 
is the Coulomb potential; however, the nonabelian nature of QCD causes
this instantaneous interaction to
depend on the gauge field, making it
intrinsically nonperturbative for large fields.
The restriction of the transverse gluon field to the fundamental modular 
region formally makes the Coulomb potential well defined.  
It also implies that the Faddeev-Popov (FP) operator 
which enters in the Coulomb potential is positive
definite~\cite{Zwanziger:1998ez,Cucchieri:1997ja}.
A consequence of this
is that one may employ the variational principle to build nonperturbative models of the 
QCD ground state.  This is a crucial step with many phenomenological 
repercussions in the methodology we will be advocating.  As we shall
demonstrate, the 
Fock space which is built on our variational vacuum consists of quasiparticles 
-- constituent quarks and gluons. These degrees of freedom obey dispersion
relations with infrared divergences due to the long-range instantaneous Coulomb
interaction of the bare partons with the 
mean field vacuum. This interaction makes colored objects infinitely
heavy thus effectively removing them from the physical spectrum.
However, color neutral states remain physical because the 
infrared singularities responsible for the large self-energies 
are canceled by infrared divergences responsible for the long-range
forces between the constituents.

Constructing a quasiparticle 
basis is a nontrivial step which requires a nonperturbative
treatment of QCD and, more directly, the QCD vacuum. 
We will show that it is possible to construct such a basis in a 
self-consistent manner
by  coupling a specific variational ansatz for the vacuum with the 
instantaneous
interaction between color charges. The end results are explicit 
expressions for
the Wilson confinement interaction, the spectrum of the quasiparticles, and the
structure of the QCD vacuum. The resulting Fock space and effective Hamiltonian
provide an ideal starting point for the  examination of the bound state problem
in QCD and provide a direct link between QCD and the phenomenological constituent
quark model.

A simplified version of this program has been investigated by the
authors and others
before~\cite{Finger:1982gm,Adler:1984ri,LeYaouanc:1985dr,lisbon,Szczepaniak:2000bi,Szczepaniak:2000uf,Szczepaniak:1997tk,Szczepaniak:1997gb,glueballs}.
In several of these studies the nonabelian Coulomb interaction was replaced by 
an effective potential between color charges, leading to a relatively simple 
many-body Hamiltonian with two-body interactions between
constituents. The phenomenology of this approach has proven quite 
successful.
In Ref.\cite{Szczepaniak:2000uf}
we have extended this simple approximation and treated the Coulomb
kernel in a self-consistent way by considering 
the effect of resummation of a class of ladder diagrams. These
diagrams originate from dressing the bare 
$\alpha_S/|\x-\y|$ Coulomb potential with transverse gluons.  
As one may expect from the discussion above, the effect of summing 
these diagrams is an enhancement of the 
Coulomb potential at large distances. Self-consistency appears in the problem
because the
strength of this enhancement is determined by the spectral properties the 
transverse gluons in the quasiparticle vacuum.

In this paper we build on these findings by constructing a fully self-consistent
set of equations which describe the gluon dispersion
relation, the effective instantaneous interaction, and the structure of the 
quasiparticle vacuum. A detailed derivation is given in Sec.~II. This section also contains a brief
review of the QCD Hamiltonian in Coulomb gauge and a discussion of the Gribov
ambiguity.
We discuss the renormalization procedure and show 
how the various counterterms in the regularized Coulomb gauge 
Hamiltonian may be constrained by physical observables. The last portion of
Sec.~II describes the variational vacuum employed in our method.
Section~III presents the solution to the coupled equations. We first
discuss the details of the renormalization procedure and present an 
approximate analytical solution which demonstrates many of the features
which emerge. This is followed by a full numerical solution and a discussion
of the effects of higher order terms.
A comparison of these results to lattice data is presented in
Sec.~IV. Section~V discusses the implications of our results for the 
 constituent
 quark model and phenomenology in general. This includes clarifying several
open issues in the CQM and extending the CQM.
A comparison to similar approaches and our conclusions are presented in Sec.~VI.

 \section{Quasiparticle Fock Space for Coulomb Gauge QCD and Confinement} 
 
One of the advantages of Coulomb gauge is that all degrees of freedom
 are physical. This makes the QCD Hamiltonian close in
spirit to quantum mechanical models of QCD, for example the constituent quark
model. The intuition gained from several decades of quark model calculations may
then be applied to the analysis of a complex and nonlinear quantum field theory.
Additional advantages of Coulomb gauge are that Gauss's law is built into the 
Hamiltonian, the norm is positive definite, and 
 no additional constraints need be imposed on Fock space.
Furthermore, retardation effects are 
 minimized for heavy quarks; thus this is a natural framework for studying
 nonrelativistic bound states, and in particular for
 identifying the physical mechanisms which drive relativistic
 corrections, {\it e.g.} the spin splittings in
heavy quarkonia.  Since chiral symmetry is
 dynamically broken this framework is also of relevance for 
  light flavors once the constituent quarks are identified with the
 quasiparticle excitations. 

The confinement phenomenon in QCD has two complementary aspects: (1)
there is a long range attractive potential between colored sources;
(2)
the gluons which mediate this force are absent from
the spectrum of physical states.  
 Thus the mechanism for confinement is not
 particularly transparent \cite{Zwanziger:1998ez} in covariant gauges.
  In Coulomb gauge, in contrast, 
 these two aspects can comfortably co-exist:
 the long range force is represented by the instantaneous 
Coulomb interaction and is enhanced as ${\bf q}^2
\to 0$, while the physical (transverse) gluon propagator
 is suppressed -- reflecting the absence of colored states in the physical 
 spectrum. 

\subsection{Coulomb Gauge Hamiltonian}

Since the Hamiltonian in Coulomb gauge may look unfamiliar to
many readers we briefly  illustrate the derivation of the classical
Hamiltonian here.


The chromoelectric field is given by
\begin{equation}
\E^a = - \dot\A^a- \bdel A^{0a} + g f^{abc}A^{0b}\A^c,
\end{equation}
and satisfies Gauss's law, 

\begin{equation}
\bdel \cdot \E^a + g f^{abc} \A^b \cdot \E^c = g \rho^a_q.
\label{gauss}
\end{equation}
Here $\rho^a_q=\psi^{\dag}( \lambda^a/2) \psi$ is the quark color
charge density. 
These equations are simplified by introducing the covariant derivative in
the adjoint representation,

\begin{equation}
\D^{ab} = \delta^{ab}\bdel +  i g  T^c_{ab} \A^c, 
\end{equation}
where $T^c$ are the adjoint representation generators, 
 $T^c_{ab} = i f^{cab}$.
Thus Eq.~(\ref{gauss}) becomes

\begin{equation}
\D^{ab}\cdot\E^b = g \rho_q^a.
\label{gauss2}
\end{equation}
If the electric field is split into transverse and longitudinal
 pieces, $\E \equiv \E_{tr} - \nabla\phi$ then Eq.~(\ref{gauss2})  yields

\begin{equation}
-(\D^{ab}\cdot \bdel) \phi = g\rho^a,
\end{equation}
 where $\rho^a = \rho^a_q + \rho^a_g$ is the full color charge
  density, with $\rho^a_g = f^{abc}\E^b_{tr}\cdot \A^c$ being the
  color charge density of transverse gluons. 
   The equation of motion for the longitudinal component of the electric
  field, 

\begin{equation}
\bdel \cdot \E^a = -\bdel \cdot \D^{ab} A^{0b} = -\bdel^2 \phi^a, 
\end{equation}
leads to a constraint for the 0-th component of the vector potential
which can be formally solved. This yields 

\begin{equation}
A^{0b} =  {1 \over \bdel \cdot \D} (-\bdel^2) {1 \over \bdel \cdot \D}
 g\rho^b,
\label{A0}
\end{equation}
and
\begin{equation}
\phi^a = {g\over {\bdel \cdot \D}}\rho^a.
\end{equation}
Finally the time evolution of the vector potential is determined by
the transverse chromoelectric field through

\begin{equation}
\bPi^a \equiv -\E_{tr}^a = \dot\A^a - g (1 - \bdel^{-2} \bdel
\bdel\cdot) f^{abc} A^{0b} \A^c.
\end{equation}
After canonical quantization, the transverse field 
 $\bPi^a$ becomes the momentum conjugate to the transverse vector
 potential, $\A^a$.

Passing from the Lagrangian to the Hamiltonian yields terms proportional to
$(\bdel\phi)^2$ from the longitudinal components of the 
 chromoelectric field in $\E^2$, terms proportional to 
 $g \rho_q A^0$ from the quark gluon vertex, $g{\bar \psi}\gamma^0
 A^{0a} \lambda^a/2\psi$  
and terms proportional to $g \Pi^a\cdot \A^b A^{0c} f^{abc}$ from  
 the  $\E_{tr}\cdot \dot{\A_{tr}}$ pieces of $\E^2$. 
Combining all these contributions and substituting the expression for 
  $A^0$ from Eq.~(\ref{A0}) results in the instantaneous
nonabelian Coulomb interaction,

\begin{equation}
H_C = {1\over 2} \int d^3x d^3y\thinspace
\rho^a({\bf x}) K_{ab}({\bf x},{\bf y};\A)  \rho^b({\bf y}),
\label{hc}
\end{equation}
where  
\begin{equation}
K_{ab}({\bf x},{\bf y};\A) \equiv \langle{\bf x},a|
 { g \over { \bnabla\cdot {\bf D}}}(-\bnabla^2)
 { g \over { \bnabla\cdot{\bf D}}}|{\bf y},b\rangle,
\label{ck}
\end{equation}
and $\rho^a$ is the full color charge density as derived above,

\begin{equation}
\rho^a({\bf x}) = \rho^a_g(\x) + \rho^a_q(\x) = 
 f^{abc} {\bf A}^b({\bf x}) \cdot {\bf \Pi}^c({\bf x}) + 
 \psi^{\dag}(\x){\lambda^a\over 2}\psi(\x).
\label{rho}
\end{equation}
The transverse conjugate gluon momenta $\bPi^a$ satisfy 
\begin{equation}
\left[ A^{a,i}(\x), \Pi^{b,j}(\y) \right] = i\delta^{ab}
\left( \delta^{ij} - {{ \nabla^i
\nabla^j} \over {\bbox{\nabla}^2} }\right)\delta(\x-\y) \equiv 
i\delta^{ab} \delta_T(\hat\bnabla)_{ij}\delta(\x-\y).
\end{equation}

Following Lee~\cite{Lee:1981mf}, we use the notation $\langle \x,a|\cdots 
 |\y,b\rangle$ to denote kernels of integral operators, 
\begin{equation}
\langle \x, a|\D|\y, b\rangle = \left[\delta^{ab}\bnabla_{\x} 
  + g f^{acb}\A^c(\x)  \right]\delta^3(\x - \y).
\end{equation}

In the abelian limit 
 ${\bf D} \to \bnabla$, $K \to -g^2 \langle \x,a|1/\bnabla^2 
 |\y,b\rangle = g^2 \delta^{ab}/4\pi|\x-\y|$  and
 the QED Coulomb interaction is recovered.

A rigorous derivation of the nonabelian, quantum Coulomb gauge
Hamiltonian was given by Schwinger\cite{Schwinger} and Christ and Lee\cite{Christ:1980ku}.
Zwanziger has shown how to derive the
Coulomb gauge Hamiltonian with a lattice regularization\cite{Zwanziger:1998ez}.
The quantum Hamiltonian may be derived by transforming the canonical $A^0 = 0$ 
Hamiltonian to Coulomb gauge.  The $A^0=0$ Hamiltonian corresponds to 
 `Cartesian'
coordinates in a flat gauge manifold, the subsequent restriction to  Coulomb gauge 
induces curvature in the gauge manifold and therefore introduces a nontrivial 
metric.
Christ and Lee have shown that the
measure associated with this metric is proportional to the Faddeev-Popov determinant

\begin{equation}
{\cal J} = {\rm det}(\nabla\cdot D).
\end{equation}
Furthermore, the Hamiltonian contains factors
of ${\cal J}$ which are analogous to the Laplace-Beltrami operator 
 induced when one first quantizes in curvilinear
 coordinates. 
The Faddeev-Popov determinant may be removed from the measure 
by working with the modified Hamiltonian

\begin{equation}
H \to  {\cal J}^{1/2} H {\cal J}^{-1/2},
\end{equation} 
which is hermitian with respect to $( \Phi|\Psi ) = \int {\cal D\A}
\Phi^*(\A) \Psi(\A)$.  Thus the final form for the
QCD Hamiltonian in Coulomb gauge is

\begin{equation}
H = H_q + H_g + H_{qg} + H_C \label{TDLeeham},
\end{equation}
where

\begin{equation}
H_q = \int d\x \psi^\dagger\left( -i \bbox{\alpha}\cdot\bdel
 + \beta m\right) \psi,
\end{equation}


\begin{equation}
H_g = {1\over 2}\int d\x \left( {\cal J}^{-1/2}\bPi {\cal J} 
\cdot \bPi {\cal J}^{-1/2} + \B\cdot\B\right),
\end{equation}

\begin{equation}
H_{qg} = -g \int d\x \psi^\dagger \bbox{\alpha}\cdot \A \psi,
\label{hqg}
\end{equation}

\noindent
and


\begin{equation}
H_C = {1\over 2}\int d\x d\y {\cal J}^{-1/2} \rho^a(x) 
 {\cal J}^{1/2} K_{ab}(\x,\y;\A) {\cal J}^{1/2} \rho^b(y) {\cal J}^{-1/2},
\label{hc2}
\end{equation}

\noindent
In order to compare with the covariant Feynman
rules and the canonical path integral formalism, it is convenient to Weyl order
the operators (we note that Weyl ordering is the operator ordering which
corresponds to path integral quantization with midpoint discretization).
This leads to the Schwinger-Christ-Lee  terms, $V_1$ and
$V_2$~\cite{Lee:1981mf}.  Here we will keep the original ordering of 
Eqs.~(\ref{TDLeeham}-21) so that no explicit $V_1$ and $V_2$ terms are present.

\subsection{The Gribov Ambiguity}

As detailed by Zwanziger \cite{Zwanziger-renor}, not only is
the Hamiltonian renormalizable in Coulomb gauge but the Gribov problem
can also be resolved \cite{Zwanziger:1998ez}.
The essence of the Gribov problem is that the condition $\bdel\cdot\A=0$
does not uniquely fix the gauge in non-Abelian
gauge theories; in general there are many copies of 
gauge field configurations, all with the same divergence,
which are related by gauge transformations.  
Alternatively, 
the canonical transformation to Coulomb gauge is not singular so long
as ${\rm det}(\nabla\cdot D) \neq 0$. But Gribov has shown that
 large 
 gauge configurations exist such that this condition does not hold.
As the
true physical configuration space of a gauge theory is the set of
gauge potentials modulo local gauge transformations, one must select a
single representative from each set of gauge-equivalent
configurations.  The resulting subset of
independent field configurations is known as the fundamental modular
region (FMR). 

A convenient characterization of the FMR is given by the
``minimal'' Coulomb gauge, obtained by minimizing a suitably chosen
functional over gauge orbits.  This functional is defined as 
\begin{equation}
F_\A [g] = {\rm Tr}\int d^3x (\A^g)^2\; ,
\end{equation}
where $g(\x)$ is a gauge transformation and 
$\A^g = g\A g^\dagger - g\bdel g^\dagger$. 
A simple calculation show that fields in the FMR are transverse. 
Alternatively, Zwanziger has demonstrated that Gribov copies may be removed by 
imposing the constraint $\langle G\rangle / {\cal V} = 0$ (called the
horizon condition) and argued that in the infinite volume limit
imposing the horizon condition enables one to remove the 
 direct restriction on the fields. 
Here $G$ is the `horizon term' given by
\begin{equation}
G = \int d\x\,d\y\, \D^{ca}(x) \cdot \langle \x a|{-1\over \bdel\cdot \D}| \y b\rangle\cdot  \D^{bc}(y) + (N_c^2-1) {\cal V}.
\label{horizon}
\end{equation}

In this paper we follow a third approach.
Because the Faddeev-Popov operator is positive semi-definite for
fields in the FMR,
we expand it in a power series over field variables and evaluate
matrix elements by integrating over all fields. 
 This is justified as long as the expectation value of the
 Faddeev-Popov operator does not change sign. We discuss under
 what conditions this procedure is consistent with the horizon condition in Sec. IV.B.
 
\subsection{Regularization and Renormalization}

To properly define the Hamiltonian a cutoff must be introduced to
regularize
ultraviolet divergences. This can be done, for example, by point splitting 
products of fields in the 
Hamiltonian.
  A simpler regularization procedure, adopted here, is to smear the
fields. The induced nonlocalities are removed as the cutoff is
taken to infinity.
Since in the numerical studies to follow we will be working with
renormalized quantities only (which are cutoff independent), 
we will explicitly remove the regulator making 
details of the regularization irrelevant.

Counterterms need to be added to the canonical Hamiltonian to ensure
that a cutoff independent spectrum is produced, 
 \begin{equation}
 H \to H(\Lambda) \to H(\Lambda) + \delta H(\Lambda). 
 \end{equation}
In this paper we concentrate on the pure glue sector with at most
static quarks, and therefore we will ignore the 
part of the Hamiltonian involving momentum or spin of the quarks. 
 In the gluon sector, the presence 
 of the cutoff leads to a single relevant operator (an operator whose 
 canonical dimension is less then four). Thus $\delta H(\Lambda)$ contains 
 a term 
\begin{equation}
\delta H(\Lambda) =  \Lambda^2 {{Z_m(\Lambda)}\over 2} \int d\x \left[\A^a(\x)^2\right]_\Lambda + \cdots 
\end{equation}
where  $Z_m(\Lambda)$ is  a dimensionless 
constant, and the notation $[\cdots]_\Lambda$ represents the effect of 
regularization.
For all marginal dimension four operators 
present in the canonical Hamiltonian there will be corresponding operators 
in $\delta H(\Lambda)$ and the combination of the two leads to a 
Hamiltonian in which canonical terms are multiplied by $\Lambda$-dependent 
renormalization constants. For example, 
 \begin{equation}
 \int d\x \left[{\bbox \Pi}^a(\x)^2\right]_\Lambda +
  \delta \int d\x \left[{\bbox \Pi}^a(\x)^2\right]_\Lambda  \equiv
  Z_\Pi(\Lambda) \int d\x \left[{\bbox \Pi}^a(\x)^2 \right]_\Lambda .
\end{equation}
The full regularized Hamiltonian with counterterms is then given by 
\begin{eqnarray}
H & = & {{Z_\Pi(\Lambda)}
\over 2}\int d\x \left[\bbox{\Pi}^a(\x)^2\right]_\Lambda
  + {{Z_A(\Lambda)}\over 2} \int d\x \left[ {\bf B}^a(\x)^2 \right]_\Lambda 
   + \Lambda^2 {{Z_m(\Lambda)}\over 2} \int \left[ \A^a(\x)^2 \right]_\Lambda
\nonumber \\
 & & + {{Z_K(\Lambda)} \over 2} \int d\x d\y \left[\rho^a(\x) K_{ab}(\x,\y;\A) \rho^b(\y)
 \right]_\Lambda + \ldots 
 \label{hamren}
\end{eqnarray}
The ellipsis stands for higher order terms induced by expanding the
  modified conjugate
 momenta ${\cal J}^{-1/2}\bPi{\cal J}^{1/2}$ in terms 
 of gauge potentials. The effect of these terms will be discussed in
 Sec.~III.F.

At this stage we should in principle allow for every composite operator of 
dimension $d$ appearing in the Hamiltonian  to be multiplied by 
a renormalization factor $Z(\Lambda)\Lambda^{4-d}$ with $Z$ being
dimensionless and also allow for the coupling constant to be $\Lambda$
dependent $g \to Z_g(\Lambda)g \equiv g(\Lambda)$. For example, as
discussed earlier,  if the 
 fields are in the FMR the Coulomb kernel may be expanded in a power series 
 in $g\A$, and the order $n$ contribution would be proportional to 
\begin{equation}
 Z_n(\Lambda) \left[{{ig(\Lambda)}\over {\bnabla^2}} \A^c 
 T^c \cdot \bnabla\right]^n_\Lambda.
\label{vertexp}
\end{equation}
 Here  $Z_n(\Lambda)$ is the $n$-th order 
 triple gluon vertex (two Coulomb and one transverse) 
 renormalization constant and $g(\Lambda)$ is the 
renormalized coupling. As we will show in Sec.~III.F such vertices 
 are UV finite which implies $Z_i(\Lambda) = 1$. The contribution 
from the Coulomb kernel to the Hamiltonian can therefore be 
written in terms of only two renormalization  constants $Z_K(\Lambda)$  
and $Z_m(\Lambda)$ (and implicitly $g(\Lambda)$) as in Eq.~(\ref{hamren}).

As mentioned above, the $\Lambda$ dependence of all renormalization constants
has to be adjusted in such a way that $H$ leads to a
$\Lambda$-independent spectrum. This implies that 
the renormalization group equations may be 
determined nonperturbatively from the spectrum of $H$. 
Furthermore in order for this Hamiltonian to be consistent with QCD
(in the chiral limit) all renormalization constants $Z_i(\Lambda)$ cannot
depend on
$\Lambda$ in an arbitrary way, but instead should depend on the
scale through the coupling $g(\Lambda)$.
The renormalization group equations will be discussed in Sec.~III.C.

\subsection{Vacuum Structure}

The eigenstates of the Hamiltonian can,
in principle,  be expanded in an arbitrarily chosen complete basis which spans 
Fock space.  One choice would be to use the perturbative basis which 
diagonalizes the free Hamiltonian  $H(g=0)$. However, one expects
the description of any hadronic bound state would be very 
complicated in this basis.
Alternatively, the phenomenologically successful constituent quark model 
indicates that hadronic wavefunctions may saturate quickly with only a few
Fock space states provided these states are constructed from
constituent (quasiparticle) quarks. This 
strongly suggests that a basis which incorporates the effects of
spontaneous chiral symmetry breaking would be more efficient for describing 
hadrons and their interactions.

We expect a similar scenario to apply to the gluon sector. 
In a given hadronic state there is a large probability of finding a component
with a large number of bare, massless transverse gluons, but 
the expansion of a hadronic state may be significantly simplified 
in a transformed Fock space which is constructed from quasiparticle (massive 
 constituent) gluons.  We follow this intuition by constructing 
a vacuum upon which the quasiparticle basis is built with a 
functional Gaussian ansatz\cite{schiff}, 
 
\begin{equation}
\Psi_0[\A] = \langle \A | \omega \rangle =  \exp\left[-{1\over 2} \int \dk
   \A^a(\k)\,\omega(k)\A^a(-\k) \right].
\label{varvac}
\end{equation}
It may be shown\cite{Barnes:1980cd} that this ansatz sums all diagrams with nonoverlapping
divergences. 
Note that the perturbative vacuum is obtained when $\omega = |\k|$. The trial
function is obtained by minimizing the vacuum energy density

\begin{equation}
  {{\delta}\over {\delta \omega}} \langle \omega |H |\omega \rangle  = 0. 
  \label{vev}
\end{equation}
The vacuum state obtained from this procedure is denoted $|\omega\rangle$.
We refer to 
 $\omega$ as the gap function since it is also 
 responsible for lifting the single particle gluon 
energy beyond its perturbative value (see Fig. 5 below).

This procedure is formally equivalent to the Hartree-Fock-Bogoliubov
approximation, therefore one may also determine $\omega$ with a suitably chosen
canonical transformation. 
Perturbative gluon creation and annihilation operators are introduced 
in the standard way, 
\begin{eqnarray}
   \A^c(\x) & = & \int \dk {1\over {\sqrt{2k} }} 
   \left[ \bbox{\epsilon}(\k,\lambda) a(\k,\lambda,c) 
 +   \bbox{\epsilon}^{*}(\k,\lambda) a^{\dag}(-\k,\lambda,c)
    \right]e^{i\k\cdot\x},
   \nonumber \\
    \bbox{\Pi}^c(\x) & = & -i \int \dk \sqrt{ k \over 2}  
   \left[ \bbox{\epsilon}(\k,\lambda) a(\k,\lambda,c) 
  - \bbox{\epsilon}^{*}(\k,\lambda) a^{\dag}(-\k,\lambda,c)   \right]e^{i\k\cdot\x},
   \end{eqnarray}
with the perturbative vacuum satisfying, $a(\k,\lambda,c)|\omega(k)=k\rangle 
 = 0$. The canonical 
transformation is determined by requiring that 
the vacuum ansatz satisfies $\alpha(\k,\lambda,c)|\omega\rangle = 0$, where 
the quasiparticle operators $\alpha,\alpha^{\dag}$ are related to the fields by 
    \begin{eqnarray}
   \A^c(\x) & = & \int \dk {1\over {\sqrt{2\omega(k)} }} 
   \left[ \bbox{\epsilon}(\k,\lambda) \alpha(\k,\lambda,c) 
 +   \bbox{\epsilon}^{*}(\k,\lambda) \alpha^{\dag}(-\k,\lambda,c)   \right]e^{i\k\cdot\x},
   \nonumber \\
    \bbox{\Pi}^c(\x) & = & -i \int \dk \sqrt{ {\omega(k)} \over 2}  
   \left[ \bbox{\epsilon}(\k,\lambda) \alpha(\k,\lambda,c) 
  - \bbox{\epsilon}^{*}(\k,\lambda) \alpha^{\dag}(-\k,\lambda,c)   \right]e^{i\k\cdot\x}.
   \end{eqnarray}  
The condition that emerges for 
  $\omega(k)$ from Eq.~(\ref{vev}) is identical to the condition that 
 there are no $\alpha^{\dag} \alpha^{\dag}$ or $\alpha \alpha$
 operators in the full Hamiltonian. 

\subsection{Self-Consistent Gap Equations}

The form of the QCD Hamiltonian in Coulomb gauge induces a crucial complication in the
evaluation of the ground state energy density. This is because the interaction
potential itself depends on the choice of the 
 vacuum: the kernel $K$ (Eq. 11) depends on the vector
fields which depend on the gap function (Eq. 32). 
Thus the gap function is
actually determined by a set of coupled  
equations which describe the vacuum energy density and the 
interactions which are used to obtain this energy density. 
This subsection describes how these equations are obtained;
the solution is presented in the next section.

The first step is the evaluation of the Coulomb kernel, Eq.~(\ref{ck}). 
This is greatly
simplified with the aid of the Swift equation\cite{Swift:1988za}:

\begin{equation}
K_{ab}(\x,\y;\A)_\Lambda, \label{d-k} = 
g^2(\Lambda) {d \over {dg(\Lambda)}} \langle \x,a|{ {g(\Lambda)}\over {\bnabla \cdot D}} 
 |\x,b\rangle.
\label{swift}
\end{equation}
The subscript $\Lambda$ refers to the regularization of fields
operators in the Coulomb kernel.  Thus one need only evaluate the 
Faddeev-Popov operator $g/\nabla\cdot \D$ to obtain the
full instantaneous Coulomb kernel.
 This can be done by expanding the Faddeev-Popov operator in powers of 
 $g\A$ and taking the appropriate contractions of the gluon field.
 The expansion is justified as long as the fields are restricted to the 
 fundamental modular region. In the infinite volume limit, this restriction is
not expected to affect field contractions\cite{Cucchieri:1997ja} as
 long as the expectation value of the horizon term vanishes. 
Thus the following expressions  may be used,
  
 \begin{eqnarray}
 \langle \omega|\left[\A^a(\x) \A^b(\y)\right]_\Lambda
 | \omega \rangle & = & {\delta_{ab} \over 2}\int^\Lambda dk 
  {{\delta_T(\hat \k)} \over {\omega(\k;\Lambda)} } e^{i\k\cdot(\x-\y)},
 \nonumber \\
  \langle \omega|\left[\bbox{\Pi}^a(\x) \bbox{\Pi}^b(\y)\right]_\Lambda
 | \omega \rangle & = & {\delta_{ab}\over 2} \int^\Lambda dk 
  \delta_T(\hat \k) \omega(k;\Lambda) e^{i\k\cdot(\x-\y)},
 \nonumber \\
  \langle \omega|\left[\A^a(\x)\bbox{\Pi}^b(\y)\right]_\Lambda
 | \omega \rangle & = &  -
   \langle \omega|\left[\bbox{\Pi}^a(\x) \A^b(\y)\right]_\Lambda
 | \omega \rangle  =  i {\delta_{ab}\over 2} \int^\Lambda dk 
  \delta_T(\hat \k) e^{i\k\cdot(\x-\y)}.
\end{eqnarray}
We have temporarily allowed for $\Lambda$-dependence in the gap function. This is
discussed in more detail in Sec. III.A.

The expansion of the Faddeev-Popov operator is given by

\begin{eqnarray} 
& & \langle \x a|  { {g(\Lambda)}\over {\bnabla \cdot \D}}
|\y,b\rangle_\Lambda  
=    D^{(0)}(\x,\y;\Lambda)\delta_{ab} +  \sum_{c_1,i_1}\int d\z_1 
D^{(1) c_1}_{i_1}(\x,\y,\z_1;\Lambda)_{ab} :A^{c_1,i_1}(\z_1):_\Lambda 
\nonumber \\ & & + \ldots \sum_{c_1\cdots c_n}\sum_{i_1\cdots i_n} 
\int d\z_1\cdots d\z_n  
 D^{(n) c_1\cdots c_n}_{i_1\cdots i_n}(\x,\y,\z_1,\cdots \z_n;\Lambda)_{ab}
 :A^{c_1,i_1}(\z_1)\cdots  A^{c_n,i_n}(\z_n):_\Lambda 
  + \ldots, \nonumber \\
  \label{dfull}
\end{eqnarray}
where $::$ stands for normal ordering with respect to
$|\omega\rangle$
and $c_n$ and $i_n$ refer to color and spatial components of the gluon
field respectively. 
Here $D^{(0)}$ stands for the expectation value (VEV) of the Faddeev-Popov
 operator in the ansatz vacuum, 
\begin{equation}
D^{(0)}_{ab}(\x,\y;\Lambda) = \langle \omega |\langle \x a|{ {g(\Lambda)}
\over {\bnabla \cdot\D}}
 |\y,b\rangle |\omega  \rangle_\Lambda
\end{equation}

An operator expansion of the Coulomb kernel may be defined in a similar manner

\begin{eqnarray}
& & K_{ab}(\x,\y;\A) =
  K^{(0)}(\x,\y;\Lambda)\delta_{ab} +  \sum_{c_1,i_1}\int d\z_1 
K^{(1) c_1}_{i_1}(\x,\y,\z_1;\Lambda)_{ab} :A^{c_1,i_1}(\z_1):_\Lambda 
\nonumber \\ & & + \ldots \sum_{c_1\cdots c_n}\sum_{i_1\cdots i_n} 
\int d\z_1\cdots d\z_n  
 K^{(n) c_1\cdots c_n}_{i_1\cdots i_n}(\x,\y,\z_1,\cdots \z_n;\Lambda)_{ab}
 :A^{c_1,i_1}(\z_1)\cdots A^{c_n,i_n}(\z_n):_\Lambda 
  + \ldots \nonumber \\
\label{Kfull}
\end{eqnarray}

The equation for the VEV of the FP operator is most easily expressed in terms
 of its Fourier transform which we write as

\begin{equation}
(2\pi)^3\delta(\P)
 {{d(\k;\Lambda)}\over {\k^2}}\delta_{ab} \equiv -\int d\x d\y 
   D^{(0)}_{ab}(\x,\y;\Lambda) 
e^{i\k \cdot(\x-\y)} 
 e^{i\P\cdot { {\x + \y} \over 2} } . \label{ddef}
\end{equation}
The amplitudes $D^{(n)}$ which multiply a product of $n$ gluon fields 
can be written in terms of $D^{(0)}$ and a set of vertex functions, 
$\Gamma^{(n)}$.  
To do this we first
define the Fourier transform of the $D^{(n)}$ via

\begin{eqnarray}
(2\pi)^3\delta(\k-\sum_{i=1}^n \q_i - \l)
d^{(n) c_1\cdots c_n}_{i_1\cdots,i_n}(\k,\q_1,\cdots,\q_n,\l;\Lambda)  
 & & \equiv - \int d\x d\y d\z_1 \cdots d\z_n 
e^{-i\k\cdot\x + i \l\cdot\y  + i\sum_{i=1}^n \q_i \cdot \z_i }
  \nonumber \\ 
& & \times D^{(n) c_1\cdots c_n}_{i_1\cdots,i_n}(\x,\z_1,\cdots,\z_n,\y;\Lambda) 
\label{ddefn}
\end{eqnarray}

Next we define the full transverse gluon-Coulomb vertex as $\Gamma^c_i(\k,\q,\p)$.
The Dyson equation for the full vertex is illustrated in 
Fig. 1 and is given by

\begin{eqnarray}
& & \Gamma^{c}_i(\k,\q,\k-\q;\Lambda) =  Z_1(\Lambda) T^c k^i 
 +  
  \sum_{c_1}\sum_{i_1}  \int^\Lambda {{d\l}\over {(2\pi)^3}}{1\over {2\omega(l;\Lambda)}}
 \left[  \Gamma^{c_1}_{i_1}(\k,\l,\k-\l;\Lambda)
{{d(\k-\l;\Lambda)}\over {(\k-\l)^2}} \right.\nonumber \\
& & \times \left. \Gamma^{c}_{i}(\k-\l,\q,
 \k-\l-\q;\Lambda) 
 {{d(\k-\l-\q;\Lambda)}\over {
(\k-\l-\q)^2}}
  \Gamma^{c_1}_{i_1}(\k-\l-\q,\l,\k-\q;\Lambda)
\right]. \label{gamma1}
\end{eqnarray}

\begin{figure}[hbp]
\hbox to \hsize{\hss\psfig{figure=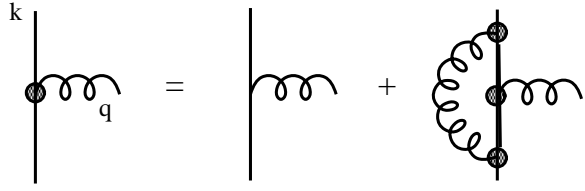,width=3.5in,angle=0}\hss}
\vspace{0.5cm}
\caption{Dyson equation of the Coulomb-transverse gluon vertex. 
The thick line
represents the full FP function $d(\k)$. 
The hatched circle represents the full vertex
$\Gamma^c_i$. The
gluon line is the gluonic quasiparticle. All external propagators are truncated.}
\label{Fig1}
\end{figure}

In the planar approximation higher order vertex functions
$\Gamma^{(n)}$ satisfy the following Dyson equation,
 
\begin{eqnarray}
& & \Gamma^{(n);c_1\cdots c_n}_{i_1\cdots
  i_n}(\k,\q_1,\cdots,\q_n,\k-\sum_{i=1}^n \q_i)
 = 
\tilde{\Gamma}^{(n);c_1\cdots c_n}_{i_1\cdots i_n}
(\k,\q_1,\cdots,\q_n,\k-\sum_{i=1}^n \q_i) \nonumber \\
& &  + \sum_{c_0}\sum_{i_0}  \int {{d\q_0}\over {(2\pi)^3}}
 {1\over {2\omega(q_0)}}
 \left[  \Gamma^{c_0}_{i_0}(\k,\q_0,\k-\q_0)
{{d(\k-\q_0)}\over {(\k-\q_0)^2}}
\tilde{\Gamma}^{(n);c_1\cdots c_n}_{i_1\cdots i_n}(\k-\q_0,\q_1\cdots
\q_n,
 \k-\q_0-\sum_{i=1}^n\q_i) \right. \nonumber \\
& & \times \left. {{d(\k-\q_0-\sum_{i=1}^n\q_n)}\over {
(\k-\q_0-\sum_{i=1}^n\q_n)^2
}}
  \Gamma^{c_0}_{i_0}(\k-\q_0-\sum_{i=1}^n\q_n,\q_0,\k-\sum_{i=1}^n\q_n)
\right],  \label{gamman}
\end{eqnarray}
where we have introduced the following quantity:

 \begin{eqnarray}
& &\tilde{\Gamma}^{(n);c_1\cdots c_n}_{i_1\cdots i_n}
(\k,\q_1,\cdots,\q_n,\k-\sum_{i=1}^n \q_i) \equiv \sum_{c_0} \sum_{i_0}
 \int {{d\q_0}\over {(2\pi)^3}}{1\over {2\omega(q_0)}}
 \left[  \Gamma^{c_0}_{i_0}(\k,\q_0,\k-\q_0)
{{d(\k-\q_0)}\over {(\k-\q_0)^2}} \right. \nonumber \\
& & \times \left. \Gamma^{c_1}_{i_1}(\k-\q_0,\q_1,\k-\q_0-\q_1) {{
d(\k-\q_0-\q_1)}\over {(\k-\q_0-\q_1)^2}}\times \right. \nonumber \\ 
& & \left. \cdots 
\times \Gamma^{c_n}_{i_n}(\k-\q_0-\sum_{i=1}^{n-1}\q_i,\q_n,\k-\q_0-\sum_{i=1}^n\q_i)
{{d(\k-\q_0-\sum_{i=1}^n\q_n)}\over {
(\k-\q_0-\sum_{i=1}^n\q_n)^2}}
\right. \nonumber \\ 
& & \times \left. \Gamma^{c_0}_{i_0}(\k-\q_0-\sum_{i=1}^n\q_n,\q_0,\k-\sum_{i=1}^n\q_n)
\right]. \nonumber \\
\label{vertex1}
\end{eqnarray}

\begin{figure}[hbp]
\hbox to \hsize{\hss\psfig{figure=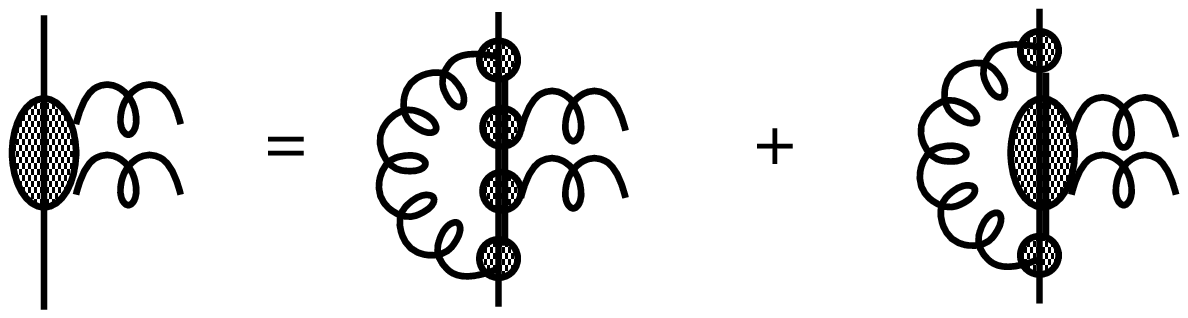,width=3.5in,angle=0}\hss}
\vspace{0.5cm}
\caption{Dyson equation for $\Gamma^{(2)}$. Symbols are as in Fig. 1.}
\label{Fig1b}
\end{figure}

The equation for $\Gamma^{(2)}$ is shown in Fig. 2.

Finally, we are able to write the coefficients of the operator product expansion of the
Faddeev-Popov operator as 

\begin{equation}
d^{(1)}_{c;i}(\k,\q,\k-\q) = d(\k) \Gamma^c_i(\k,\q,\k-\q) d(\q)
\end{equation}
and
\begin{eqnarray}
d^{(2)}_{c_1,c_2;i_1,i_2}(\k,\q_1,\q_2,\k-\q_1-\q_2)& = & 
 d(\k) \Gamma^{c_1}(\k,\q_1,\k-\q_1) d(\k-\q_1)\Gamma^{c_2}(\k-\q_1,\q_2,\k-\q_1
-\q_2) \nonumber \\
 &+&  d(\k)\Gamma^{(2) c_1,c_2}_{i_1,i_2}(\k,\q_1,\q_2,\k-\q_1-\q_2) 
d(\k-\q_1-q_2), \nonumber \\
\end{eqnarray}
and similarly for higher orders. Before
renormalization, these amplitudes are functions of the cutoff. 
In the planar approximation the VEV of the Faddeev-Popov operator, 
 $d(k;\Lambda) = d^{(0)}(\k,\k;\Lambda)$ defined in Eq.~(\ref{ddef})
 satisfies, 
\begin{equation}
d(k;\Lambda) = { {g(\Lambda)} \over {1 - g(\Lambda) I[d,\omega]}}, 
\label{dgen}
\end{equation}
where 
\begin{eqnarray}
I& & [d,\omega] = 
  \sum_{n}
{1\over {N_c^2-1}}Tr \sum_{c_1 \cdots c_n}\sum_{i_1\cdots i_n} {1\over {\k^2}}
\int^\Lambda {{d\q_1}\over {(2\pi)^3}}
\cdots  {{d\q_n}\over {(2\pi)^3}}
 {1\over {2\omega(q_1;\Lambda)\times \cdots \times 2\omega(q_n;\Lambda)
}}  \nonumber \\
& & \times \Gamma^{(n) c_1\cdots c_n}_{i_1\cdots i_n }(\k; 
 \q_i,\cdots,\q_n;\k-\sum_{i=1}^n\q_i) {{d(\k-\sum_{i=1}^n \q_i;\Lambda)}
 \over  {(\k-\sum_{i=1}^n \q_i)^2}}
 \Gamma^{(n) c_1\cdots c_n}_{i_1\cdots i_n}
 (\k-\sum_{i=1}^n\q_i;\q_1,\cdots,\q_n;\k).  \nonumber \\ 
\end{eqnarray}
The trace is taken over the implicit  color indices of the vertex
functions, $\Gamma^{(n)} = \Gamma^{(n)}_{ab}$, which also absorb the renormalization 
constants $Z_i(\Lambda)$ of Eq.~(\ref{vertexp}). 
This equation is shown in Fig.~3.

\begin{figure}[hbp]
\hbox to \hsize{\hss\psfig{figure=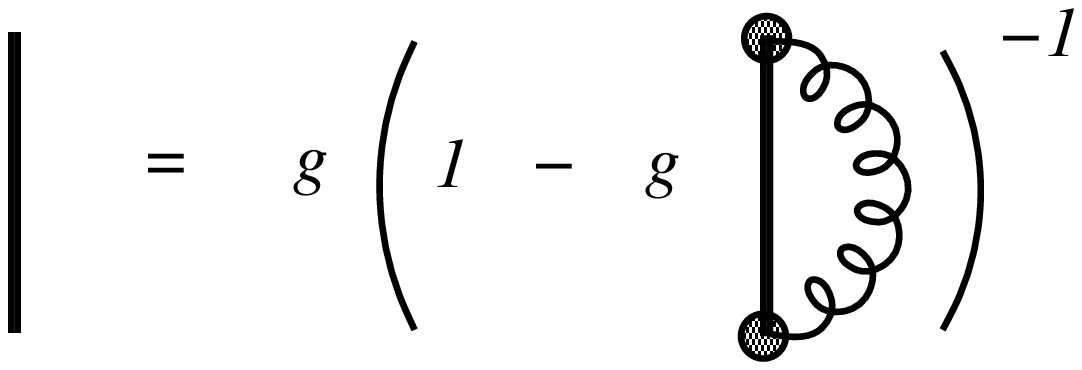,width=3.5in,angle=0}\hss}
\vspace{0.5cm}
\caption{  Dyson equation for the VEV of the Faddeev-Popov operator. See Fig.~1 for
an explanation of the symbols.}
\label{Fig2}
\end{figure}

We proceed to the evaluation of the Coulomb kernel. Following
 Swift~\cite{Swift:1988za} we define 
$f(\k;\Lambda)$ via
 \begin{equation}
(2\pi)^3\delta(\P)
 d^2(K,\Lambda) {{f(\k,\Lambda)}\over {\k^2}}\delta_{ab} \equiv  
 Z_K(\Lambda) \int d\x d\y 
   \langle \omega|K_{ab}(\x,\y) |\omega \rangle_\Lambda
e^{i\k \cdot(\x-\y)} 
 e^{i\P\cdot { {\x + \y} \over 2} }.  \label{d2f}
\end{equation}
From Eqs.~(\ref{d-k}) and (\ref{dgen}) it follows that the  
vacuum expectation value of the Coulomb kernel satisfies
\begin{equation}
f(k,\Lambda) = Z_K(\Lambda) + {d\over {d g}} I[Z_K d,\omega] .
\label{fgen}
\end{equation}
This comprises a linear integral equation which must be solved for $f$ after having
obtained $d$.
We are finally in a position to evaluate the expectation value of the energy density 
from the full Hamiltonian,
\begin{equation}
{\cal E} = {\cal E}_g + {\cal E}_m + {\cal E}_C \equiv {1\over {\cal V}(N^2_c-1)} \langle \omega |H|\omega \rangle
\end{equation}
 where the three terms represent the kinetic energy 
 (including the nonabelian portion of the ${\bf B}^2$ term), the 
 mass counterterm, and the Coulomb potential respectively. In
 particular, 

  
\begin{equation}
{\cal E}_g   =  
 {1\over 2}  \int^\Lambda \dq  
  \left[ 
  { Z_\Pi(\Lambda) \omega(q;\Lambda) } +  Z_A(\Lambda) {{q^2}\over {\omega(q;\Lambda)}} 
  \right] + 
   {g^2(\Lambda) {{N_c}\over 16} } \int^\Lambda \dq   
 \dk 
 { {(3 - (\hat \k \cdot \hat \q)^2) }\over 
 {\omega(q;\Lambda)\omega(k;\Lambda)} }
   \end{equation}
and
\begin{equation}
{\cal E}_m = {1\over 2} Z_m \Lambda^2 \int^\Lambda \dq {1\over {\omega(q;\Lambda)}} .
\end{equation}

The contribution from the Coulomb potential may be evaluated with the aid
of the operator expansion in Eq.~(\ref{Kfull}).  Recall that the products
of gluon fields in the operator expansion of the kernel 
 are normal ordered with respect to the variational vacuum. 
 Thus the maximum number of terms which contribute to the vacuum energy
energy density is determined by the number of external fields present in 
  the charge densities multiplying the kernel ({\it i.e.}, four).
 The Coulomb vacuum energy density may be thus be written as

\begin{equation}
{\cal E}_C = {\cal E}^{(0)}_C  + {\cal E}^{(2)}_C + {\cal E}^{(4)}_C.
\end{equation}
The terms ${\cal E}^{(n)}$ correspond to the vacuum expectation values of
$K^{(n)}$ contracted with the fields from the charge densities. 
For the first term one gets,  
\begin{equation}
 {\cal E}^{(0)}_C  =  {N_c\over 16} \int^\Lambda \dq \dk
   {{f(\k+\q;\Lambda)d^2(\k+\q;\Lambda)}\over {(\k+\q)^2}} 
(1 + (\hat\k\cdot\hat\q)^2 )
   \left[ {{\omega(k;\Lambda)}\over {\omega(q;\Lambda)}} 
         +{{\omega(q;\Lambda)}\over {\omega(k;\Lambda)}} - 2 \right]
\label{e0c}. 
\end{equation}
The higher order terms ${\cal E}^{(n)}_C$ are of order
$d^{(n+2)}(k;\Lambda)$. Since $d$ plays the role of the running coupling 
(see Eq.~(\ref{dgen})), we 
expect these higher order terms to give finite 
corrections to $\omega(k;\Lambda)$ 
which should be small in particular for large momenta because of the 
suppression from the running coupling. 
The effect of these higher order terms as well as vertex corrections
(c.f. Eqs.~(\ref{vertex1}) and (\ref{gamman})) and the FP determinant 
will be discussed in detail in Sec.~III.F.  

Minimizing 
${\cal E}^{(n)}_C$ with respect to $\omega$ leads to two contributions -- 
one from the explicit $\omega$ dependence (cf. Eq.~(\ref{e0c}) for  
${\cal E}^{(0)}_C$) and the other from the implicit $\omega$
dependence arising 
through the 
kernel $fd^2$. We refer to these contributions to the gap equation as 
${\cal E}^{(n),\omega}_C$  and ${\cal E}^{(n),K}_C$ respectively. 
The first of these is of order 
 ${\cal O}(d^{(n+2)}(k;\Lambda))$ and the second is 
${\cal O}(d^{(n+4)}(k;\Lambda))$.  Thus, for example 
${\cal E}^{(0),K}_C$ should be combined with other order 
${\cal O}(d^{4}(k;\Lambda))$ contributions from ${\cal E}^{(2),\omega}_C$.
Subsequent expressions for ${\cal E}_C^{(n),\omega/K}$ contain a factor
of $-2 \omega^2$ with respect to the derivatives of ${\cal E}$.
For the moment we retain only the leading ${\cal O}(d^2(k;\Lambda))$ 
contributions from ${\cal E}^{(0),\omega}_C$ in the gap equation.
Minimizing ${\cal E}$ with respect to $\omega$ leads to the following gap equation

\begin{eqnarray}
Z_\Pi^2(\Lambda)\omega^2(q;\Lambda) & = &  Z_A^2(\Lambda) q^2  + 
Z_m(\Lambda)\Lambda^2  
+ {g^2(\Lambda) {{N_c}\over 4} } \int^\Lambda \dk 
{ {(3 - (\hat \k \cdot \hat \q)^2) }\over \omega(k;\Lambda)}  +
\nonumber \\
& &  + {N_c \over 4} \int^\Lambda \dk {{f(\k+\q;\Lambda)d^2(\k+\q;\Lambda)}
\over {(\k+\q)^2}} (1 + (\hat \k \cdot \hat \q)^2)\, 
{\omega^2(k;\Lambda) - \omega^2(q;\Lambda) \over \omega(k;\Lambda)}.  \nonumber \\
\label{gapgen}
\end{eqnarray}


This completes the derivation of the leading order gap equations. To summarize, these comprise
Eq.~(\ref{dgen}) for the VEV of the FP operator 
  $d(k;\Lambda)$, Eq.~(\ref{fgen}) for the Coulomb
kernel $f(k;\Lambda)$, and Eq.~(\ref{gapgen}) for the gap function $\omega(k;\Lambda)$.

\section{Solution of the Self-Consistent Gap Equations}

Before continuing we shall briefly summarize our philosophy. 
The goal is to construct a quasiparticle Fock space which will provide a useful
starting point for the evaluation of hadronic observables. 
Quasiparticle states
 are built on a variational vacuum and reflect the propagation of these degrees of
freedom through a nontrivial background.
Of course the full Hamiltonian still contains many-body terms which mix the 
free quasiparticle states; nevertheless, the quasiparticle Fock space is
complete and at least in principle one should be able to diagonalize the full 
Hamiltonian in this basis.

When dynamical quarks and gluons are considered, one would need to 
diagonalize the 
Hamiltonian in the full Fock space. In practice, however, 
such diagonalization  is always performed in an appropriately selected
subspace  {\it e.g.} including only $|Q\bar Q\rangle$ or $|QQQ\rangle$ 
quasiparticle states. 
Such a truncation is better justified when the quasiparticles behave as 
constituent particles with average kinetic energies of several hundred MeV.
Furthermore, as discussed earlier, the quasiparticle basis diagonalizes 
the one-body part of the Hamiltonian, thus at least at the level of Tamm-Dancoff 
truncation, the quasiparticle vacuum decouples from the hadronic spectrum.

The required cut-off independence of the eigenvalues 
can be used to determine the $\Lambda$ dependence of the various 
counterterms and couplings.  At this stage, this implies that the Fock space itself should be 
cutoff-independent because, for example, the ground state energy of two static 
color sources is directly related to the expectation value of $H$ in the variational vacuum.  
We note that 
expanding the Fock space in which the Hamiltonian is being  nonperturbatively
diagonalized will add new 
counterterms to the Hamiltonian which will modify the renormalization group equations.

\subsection{Vertex Truncation}

We start by examining the renormalization group structure which follows from the 
requirement that the gluonic Fock space itself is $\Lambda$ independent. 
This implies that the Coulomb kernel, and hence $f(k;\Lambda)$ and 
$d(k;\Lambda)$, should be $\Lambda$-independent. These conditions may be imposed
through an appropriate choice of the cutoff dependence of the counterterms and coupling.

Consider first renormalizing the FP operator of Eq.~(\ref{dgen}).  In this equation $d$
is expressed in terms of the vertex functions $\Gamma^{(n)}$ and the gap
function $\omega$. Since these can be independently 
renormalized using other renormalization parameters which do not
explicitly show up in Eq.~(\ref{dgen}) ({\it i.e.} $Z_1(\Lambda),
Z_m(\Lambda)$ in Eqs.~(\ref{hamren}) and (\ref{vertexp})) 
 we can replace them by their 
renormalized, $\Lambda$-independent versions, 
$\Gamma^{(n)}(\cdots;\Lambda) \to \Gamma^{(n)}(\cdots)$ and $\omega(k;\Lambda)
\to \omega(k)$. Thus the only $\Lambda$-dependent parameter available to enforce
the cutoff independence of the FP operator is the coupling, $g(\Lambda)$. 

To determine the consequences of this observation we examine the behavior of
the vertices which appear in the equation for $d$ (\ref{dgen}).
Asymptotic freedom implies that for momenta near the UV cutoff, 
 the gap function and the renormalized
vertex functions approach their corresponding free-field values,

\begin{equation}
\lim_{k \sim \Lambda \to \infty}\omega(k) = k + {\cal O}(g^2(\Lambda))
\end{equation}
and

\begin{equation}
\Gamma^{c}_{i}(\k,\q,\k-\q) \to k^i T^c + {\cal O}(g^2(\Lambda)).
\end{equation}
For $n>1$

\begin{equation}
\Gamma^{(n)}(\k,\q_1,\cdots,\q_n,\k-\sum_{i=1}^n \q_i) \to {\cal O}(g^{n+2}(\Lambda)).
\end{equation}
 Similarly one expects that in this limit $d(k) \to {\cal O}(g(\Lambda))$. 
Thus the integral in Eq.~\ref{dgen} is 
 logarithmically divergent as $\Lambda \to \infty$. This divergence is
 absorbed by the coupling $g(\Lambda)$.
It follows  from Eq.~(\ref{gamma1}) 
that $\Gamma^c_i$ is given by an expression which is finite
as $\Lambda$ approaches infinity, thus there is no need for vertex
renormalization and one can set $Z_1(\Lambda)=1$.                            
Furthermore, the correction to the bare vertex $T^c \k$ is expected to
be of the order ${\cal O}(\langle g^2 \rangle )$ where $\langle g \rangle$
refers to an UV and IR finite integral over the running
coupling. This is due to the two Faddeev-Popov operators
$d(k)$ in Eq.~(\ref{gamma1}). Since $d(k)$ is proportional to 
 $g(\Lambda=k)$ for large $k$, the 
 renormalized FP operator can be associated with the running coupling:

\begin{equation}
\lim_{k \to \infty} d(k) \to g(\Lambda=k).
\end{equation}
From the resummation implicit in Eq.~(\ref{dgen}), and consistent with asymptotic
freedom,
the large momentum behavior of $d(k)$ will be logarithmically suppressed
with $k$. 
Furthermore, if
 $d(k)$ is less singular than $1/k$  in the infrared limit
then the integral on the right hand side of
Eq.~(\ref{gamma1}) represents a finite, higher order (in
the running QCD coupling) correction to the bare vertex.  
This is also true for the higher order irreducible vertices, $\Gamma^{(n)}$. 
From Eq.~(\ref{gamman}) it follows that these are ${\cal O}(\langle g^{n+2} \rangle)$.
This important observation will be used to truncate the gap equations in the 
next subsection.

\subsection{The Truncated and Renormalized Gap Equations}

The considerations of the previous subsection may be used to truncate the
general gap equations derived in Sec.~II. This is necessary to make
the equations tractable. The effect of neglected terms will be
discussed in Sec.~III.D . 

We start by ignoring the finite higher order corrections to the vertices
and thus take
\begin{equation}
\Gamma^c_i(\k,\q,\k-\q) = T^ck^i,
\end{equation}
and
\begin{equation}
\Gamma^{(n)} \to 0.
\end{equation}
The equation for the unrenormalized FP operator, Eq.~(\ref{dgen}), 
becomes
\begin{equation}
{1 \over {d(k;\Lambda)}} = {1\over g(\Lambda)} - N_c\int^\Lambda 
 \dq  { {1 - ({\hat \k}\cdot {\hat \q})^2} \over 
 {2\omega(q)(\k-\q)^2} } d(\k-\q;\Lambda).
\label{dtrun}
\end{equation}
One sees from this equation that in
order for $d(k;\Lambda)$ to be $\Lambda$ independent, $g(\Lambda)$ must
obey the following 
renormalization group equation 
   
\begin{equation}
 {1\over { g(\Lambda)}} = {1\over { g(\mu) }} + 
   N_c\int^\Lambda \dq  { {1 - (\hat\q\cdot \hat{\bbox{\mu}})^2} \over 
 {2\omega(q)(\q-\bbox{\mu})^2} } 
 d(\q-\bbox{\mu}).
\label{reng}
\end{equation}
Thus Eq.~(\ref{dtrun}) becomes
 
 \begin{equation}
 {1 \over {d(k)}} = {1\over d(\mu)} - N_c\int
 \dq  { {1 - ({\hat \q}\cdot {\hat \k})^2} \over 
 {2\omega(q)(\q-\k)^2} } d(\q-\k)
 + N_c\int\dq  { {1 - ({\hat \q}\cdot \hat{\bbox{\mu}})^2} \over 
 {2\omega(q)(\q-\bbox{\mu})^2} } d(\q-\bbox{\mu}).
 \label{rend}
 \end{equation}
Here the renormalized FP operator is written as $d(k;\Lambda) \to d(k)$.
Eq.~(\ref{rend}) implies   that $d(k)$ is independent of $\Lambda$ (and the
scale $\mu$), and represents the once-subtracted form of Eq.~(\ref{dtrun}).
The presence of $g(\Lambda)$ in Eq.~(\ref{dgen}) shows that 
$d(k)$ can only be determined 
up to an overall constant. Thus the equation for $d(k)$ contains a
single unknown, $d(\mu)$.

The vertex truncations and Eqs.~(\ref{fgen}) and  (\ref{rend}) imply that
the expectation value of the unrenormalized Coulomb kernel is given by 

\begin{equation}
f(k,\Lambda) = Z_K(\Lambda) +  N_c\int
 \dq  { {1 - ({\hat \q}\cdot {\hat \k})^2} \over 
 {2\omega(q)(\q-\k)^2} } d^2(\q-\k) f(\q-\k;\Lambda).
\end{equation}
The UV divergence from the integral is absorbed by 
$Z_K(\Lambda)$. Subtracting once yields 
\begin{eqnarray}
f(k) = f(\mu) + & &  N_c\int
 \dq  { {1 - ({\hat \q}\cdot {\hat \k})^2} \over 
 {2\omega(q)(\q-\k)^2} } d^2(\q-\k) f(\q-\k) \nonumber \\
  - & &  N_c\int\dq  { {1 - ({\hat \q}\cdot \hat{\bbox{\mu}})^2} \over 
 {2\omega(q)(\q-\bbox{\mu})^2} } d^2(\q-\bbox{\mu}) f(\q-\bbox{\mu}).
\label{renf}
\end{eqnarray}
Here $f(\mu)$ is another external renormalization parameter.  The renormalization
constant is given in terms of it by

\begin{equation}
Z_K(\Lambda) = f(\mu)  
 - N_c\int\dq  { {1 - ({\hat \q}\cdot \hat{\bbox{\mu}})^2} \over 
 {2\omega(q)(\q-\bbox{\mu})^2} } d^2(\q-\bbox{\mu}) f(\q-\bbox{\mu}).
\label{zk}
\end{equation}

We finally discuss renormalization of the gap equation,
Eq.~(\ref{gapgen}). In general this equation can depend on the 
three renormalization constants, $Z_A(\Lambda)$,
$Z_\Pi(\Lambda)$, $Z_m(\Lambda)$ and the renormalized coupling,
$g(\Lambda)$.
The coupling  $g(\Lambda)$ is already determined
by Eq.~(\ref{reng}).
In the UV limit the integral on the right hand side of Eq.~(\ref{gapgen})
has  in principle quadratic and logarithmic divergences. The logarithmic  divergence is present 
if the kernel $f(\k-\q)d^2(\k-\q)$ approaches a constant in the UV limit. There 
are, however,  logarithmic corrections to both $f$ and $d$ which follow from
Eqs.~(\ref{rend}) and ~(\ref{renf}) which actually protect the integral from the 
logarithmic divergence.
Thus one can immediately set $Z_A = Z_\Pi= 1$ and
absorb all possible remaining divergences (as $\Lambda \to \infty$) into 
$Z_m(\Lambda)$. 
This leaves the quadratic divergence which is eliminated
by a single subtraction, 

\begin{eqnarray}
 \omega^2(q) = q^2 - \mu^2 + \omega^2(\mu) 
    & + & {N_c \over 4}\int\dk  
{{f(\k+\q)d^2(\k+\q)}
 \over {(\k+\q)^2}}
 (1 + 
 (\hat \k \cdot \hat \q)^2)\, {\omega^2(k) - \omega^2(q)
  \over
   \omega(k)} \nonumber \\
 & - &  {Nc\over 4} \int \dk 
{{f(\k+\bbox{\mu})d^2(\k+\bbox{\mu})}
 \over {(\k+\bbox{\mu})^2}}
  (1 +  (\hat \k \cdot \hat{\bbox{\mu}})^2)
 { {\omega^2(k) - \omega^2(\mu) } \over
   \omega(k)}.
\label{renom}
\end{eqnarray}
The mass counterterm is given in terms of $\omega(\mu)$ by 

\begin{eqnarray}
Z_m(\Lambda)\Lambda^2 =& &  \omega^2(\mu) - \mu^2 - 
  {g^2(\Lambda) {{N_c}\over 4} } \int^\Lambda \dk 
 { {(3 - (\hat \k \cdot \hat \q)^2) }\over \omega(k)}
 \nonumber \\
& &  - {N_c \over 4} \int^\Lambda \dk  
{{f(\k+\bbox{\mu})d^2(\k+\bbox{\mu})}
 \over {(\k+\bbox{\mu})^2}}
 (1 + 
 (\hat \k \cdot \hat{\bbox{\mu}})^2)\, {\omega^2(k) - \omega^2(\mu)
  \over
   \omega(k)} .
\label{zm}
\end{eqnarray}

Equations (\ref{rend}), (\ref{renf}), and (\ref{renom}) form the
 renormalized coupled gap equations which represent the
leading order vacuum and quasiparticle structure of QCD in Coulomb gauge.
We proceed by examining the perturbative limit of these equations before
turning to analytical and numerical solutions. 
Subsection III.F examines corrections to the gap equations due to
 truncation to the leading terms.

\subsection{Asymptotic  Renormalization Group Equations}

We establish the relationship of the  renormalized gap equations
to standard perturbative QCD in this section.
The renormalization group
equation for the renormalized coupling, Eq.~(\ref{reng}) implies that
for large cutoffs
 
\begin{equation}
  \Lambda {{dg(\Lambda)} \over {d\Lambda}} 
   = -{{8N_C}\over {3}} {{g^2(\Lambda)d(\Lambda)}\over {(4\pi)^2}}, \label{zg}
\end{equation}
and from Eqs.~(\ref{reng}) and ~(\ref{rend}) it follows that in the limit 
 $\Lambda \to \infty$ 
\begin{equation}
  \Lambda {{dg(\Lambda)} \over {d\Lambda}} 
   = -{{8N_C}\over {3}} {{g^3(\Lambda)}\over {(4\pi)^2}} \equiv
  \beta(g(\Lambda)). \label{zg1}
\end{equation}

We call the first coefficient in the expansion of the $\beta$ function
$\bar \beta_0$. The last equation implies that

\begin{equation}
\bar\beta_0 = {8 N_c \over 3}.
\end{equation}
 Although it is tempting to compare this to the canonical perturbative
 expression of $\beta_0 = 11 N_c/3$, this is
 misleading for two reasons. 
First the coupling 
defined here corresponds to the product
of the VEV of a composite operator ({\it i.e.} the Faddeev-Popov operator)  and the QCD
coupling. Thus $\bar\beta$ will also reflect renormalization of
the FP operator. We note that this is nevertheless a sensible definition for the coupling
since it is this product which determines the strength of
the various interactions involving Coulomb gluons. 
The second reason is that we sum loops which arise from the expectation value of the
Hamiltonian and do not include those from iterating the Hamiltonian.
Iteration of the Hamiltonian involves summing over  
intermediate states. This is fine in perturbation theory, but 
because of confinement can only be justified for 
color singlets so that summation should be restricted to hadronic intermediate
states only. 
As discussed in Sec.~II.D this  may be achieved 
in bound state perturbation theory once the 
quasiparticle Fock space is specified. We will discuss the
 running coupling in more detail below. 


The expression for $Z_K$ given in Eq.~(\ref{zm}) implies that the 
renormalization group equation for $Z_K(\Lambda)$  is  
given by
\begin{equation}
\Lambda {{Z_K(\Lambda)}\over {d\Lambda}} = -{\bar\beta}_0 
{{d^2(\Lambda)f(\Lambda)}\over {(4\pi)^2}},
\end{equation}
which in the limit $\Lambda \to \infty$ leads to 
\begin{equation}
\Lambda {{dZ_K(\Lambda)}\over {d\Lambda}} = -{{8N_C}\over 3}
{{g^2(\Lambda)}\over {(4\pi)^2}}Z_K(\Lambda).
\label{zkpert}
\end{equation}
Finally, Eq.~(\ref{zm}) yields

\begin{equation}
\Lambda {{ dZ_m(\Lambda)}\over {d\Lambda}} 
 = - 2Z_m(\Lambda) - 
{{g^2(\Lambda)}\over {(4\pi)^2}}\bar\beta_0\left[ 
  2\left(1 + {{g^2(\Lambda)}\over {(4\pi)^2}}\bar\beta_0\right)
 + Z_K(\Lambda) \right].
\label{zmpert}
\end{equation}
The first term is universal and reflects the
quadratic divergence. The remainder relates to the UV behavior of the
Coulomb kernel and the quartic-gluon vertex which are both determined
by the running coupling $g(\Lambda)$.


As expected, all counterterms run as a function of a single renormalized parameter 
$g(\Lambda)$, where from Eq.~(\ref{zg1}),

\begin{equation}
g^2(\Lambda) =  {{g^2(\Lambda_0)} \over { 1 + {{\bar\beta_0}\over {(4\pi)^2}}
g^2(\Lambda_0)\log{{\Lambda^2}\over {\Lambda_0^2}} } }  
 = {{(4\pi)^2} \over {\bar\beta_0 \log{{\Lambda^2}\over {\Lambda_{QCD}^2}  } } } ,
\end{equation}
with

\begin{equation}
\Lambda_{QCD}^2 = \Lambda^2 \exp\left(-(4\pi)^2/\bar\beta_0 g^2(\Lambda)\right).
\end{equation}

Solving the renormalization group equations and substituting for $g$ yields
the following expressions for the mass and Coulomb renormalization constants

\begin{equation}
Z_K(\Lambda) = Z_K(\Lambda_1) 
\left( {{\log {{\Lambda_1}\over {\Lambda_{QCD}^2} } }
 \over  {\log {{\Lambda}\over {\Lambda_{QCD}^2} } } } \right)^{1\over 2}
 = Z_K(\Lambda_1) {{g(\Lambda)} \over {g(\Lambda_1)} } ,
\end{equation}
and
\begin{eqnarray}
\Lambda^2 Z_m(\Lambda) & = &  Z_m(\Lambda_1)\Lambda^2_1 
  - {\bar\beta_0\over {(4\pi)^2}}
  \int_{\Lambda^2_1}^{\Lambda^2}  d t 
 g^2(t) \left[ \left(1 + {{g^2(t)}\over {(4\pi)^2}}\bar\beta_0\right)
 + {1\over 2}Z_K(t) \right]. 
\end{eqnarray}     

Lastly, we examine the effective renormalized  potential 
between static color sources. This may be defined via Eqs.~(\ref{hc}) and
(\ref{d2f}) as
 
\begin{equation}
V(k) \equiv  {{f(k)d^2(k)} \over {k^2}} \equiv { {4\pi\alpha_{\rm eff}(k)}
  \over {k^2}} \label{alphaeff}. 
\end{equation} 
It is clear that it is the combination $Z_K(\Lambda)g^2(\Lambda)$ which 
is responsible for making 
$\alpha_{\rm eff}$ $\Lambda$-independent. For large  $\Lambda$ and $\Lambda_1$
one obtains
 
\begin{equation}
\alpha_{\rm eff}(\Lambda) = \alpha_{\rm eff}(\Lambda_1) {{ Z_K(\Lambda) g^2(\Lambda)
 } \over {Z_K(\Lambda_1) g^2(\Lambda_1) }} 
  = {{ \alpha_{\rm eff}(\Lambda_1) } \over { \left( 1 + \bar\beta_0 {{g^2(\Lambda_1)} \over {(4\pi)^2}} 
   \log {{\Lambda^2} \over {\Lambda^2_1}} \right)^{3\over 2} } }.
\label{77}
\end{equation}
Notice the power in the denominator which is present due to the 
 rainbow-ladder nonperturbative structure of the VEV of the Coulomb 
 operator.
 Expanding Eq.~(\ref{77})  permits a comparison to 
perturbation theory:
 
\begin{equation}
 \alpha_{\rm eff}(\Lambda) = \alpha_{\rm eff}(\Lambda_1)\left( 
 1 + {3\over 2} \bar\beta_0 {{g^2(\Lambda_1)} \over {(4\pi)^2}} 
   \log {{\Lambda^2_1} \over {\Lambda^2}} + {\cal O}(g^4) \right) .
\end{equation}
In perturbation theory (with no light quarks) the coefficients in front of 
$g^2(\Lambda_1)$ 
should be equal to $4N_C - N_C/3 = 11$ rather than $3/2\bar\beta_0 = 4N_C$.
The difference comes from
the perturbative contribution due to emission and absorption of a transverse 
gluon, which involves iterating the Coulomb-transverse gluon vertex from 
$H_C$ twice. This contribution is not present when one
takes the expectation value of the Hamiltonian as done here. However, 
 as stated earlier, perturbative 
contributions from propagating transverse gluons may be included, for
example, in bound state perturbation theory and can be systematically
included in
our approach when the Hamiltonian is diagonalized in the
quasiparticle basis. 
 It should also be noted that such
differences are of a screening nature, and thus are not expected to
spoil the confinement mechanisms coming  from summing the
Coulomb-transverse gluon interactions.

\subsection{Approximate Analytical Solution} 

In this subsection we present an approximate analytical solution
 to the truncated renormalized coupled gap equations for $d(k)$, $f(k)$ and $\omega(k)$; 
Eqs.~(\ref{rend}),~(\ref{renf}), ~(\ref{renom}), respectively. 
The approximate solution is obtained by simplifying 
 the angular part of the integrals over 3-momenta. In each case the
 angular dependence is approximated by 

\begin{equation} 
(\k - \q)^2 \to \theta(k^2 - q^2) k^2 + \theta(q^2 - k^2) q^2.
\label{ang}
\end{equation}
Next we assume  that the renormalized solution of the gap equation can be
written in the form 

\begin{equation}
\omega(k) = \theta(k - m_g) k + \theta(m_g - k) m_g.
\label{omaprox}
\end{equation}
Thus we assume that the gap function saturates to a nonzero
value at low momentum.
Once the FP operator $d(k)$ and the Coulomb kernel $f(k)$ have been obtained,
the gap equation may be solved for $\omega(k)$ and the 
consistency of the ansatz for $\omega$ may be checked.

With the aid of these approximations the equation for the running coupling
can be converted into differential form

\begin{equation}
-{{d'(k)}\over {d^2(k)}} = \left\{ \begin{array}{cc} 
{{\bar\beta_0} \over {(4\pi)^2}} \left(
 {2\over 3} {{d(k)} \over {m_g}} 
 - {1\over 3}  {k\over {m_g}} d'(k)\right) & \mbox{ for } k
 \le m_g, \\
{{\bar\beta_0} \over {(4\pi)^2}} 
  \left( 
 1 - { {m_g^2} \over {3 k^2}} \right) 
 \left( {{d(k)}\over k}   - {1\over 2} d'(k) 
  \right)  & \mbox{ for } k
 \ge m_g  \end{array} \right. .
\label{diff}
\end{equation}
For  $k\le m_g$ the solution is given by
\begin{equation}
{{d(k)^2} \over 
{ | 3 - {{5 \bar\beta_0}\over {3(4\pi)^2}} {k\over {m_g}} d^2(k)  |^{4\over 5} } }
 = {{d(\mu)^2 } \over { | 3 - {{5 \bar\beta_0}\over {3(4\pi)^2}} {{\mu}
       \over {m_g}}      d^2(\mu)  |^{4\over 5} } },
\label{klgex}
\end{equation}
which is well approximated by
\begin{equation}
d(k) = {{d(\mu)} \over { \left[1 + {{5\bar\beta_0} \over {3(4\pi)^2}} d^2(\mu) \left( 
  {{ k - \mu} \over {m_g}} \right)\right]^{1\over 2} } }.
\label{klg}
\end{equation}
This equation is trivially $\mu$-independent. 
For large momenta, $k \ge m_g$  we approximate Eq.~(\ref{diff}) by
neglecting the terms of $O(m_g^2/k^2)$. In this case the solution is given
 by 
\begin{equation}
 d(k) = {{ d(\mu)} \over {\left(1 + {\bar\beta_0 \over {(4\pi)^2} }d(\mu)^2 \log 
 {k^2 \over \mu^2} \right)^{1\over 2} }},
\label{kgm}
\end{equation}
which also is $\mu$-independent. 
Even though this solution is valid for $k >> m_g$ it may be matched 
continuously with the solution for $k_g < m_g$ if one chooses $\mu = m_g$. 
The freedom in the renormalization of $d(k)$ is now related to the choice
of the value of $d(k)$ at $k=\mu=m_g$.


It follows from Eqs.~(\ref{klgex}) and (\ref{klg}) that there is a critical value 
of  $d(m_g) = d_c = 4\pi \sqrt{3/5\bar\beta_0} \sim  3.4414$ for $N_c =3$ 
which leads to $d(k) \propto 1/\sqrt{k}$ for small $k$. Furthermore, this
is the strongest possible IR enhancement admitted by the approximate
solution.  The solution for $d$ approaches a finite value for all
other values of
$d(m_g)$ less than $d_c$. We shall see that this general behavior remains true for
the full numerical solution as well.
   
 The corresponding solution for the function $f(k)$
 follows from 
 Eq.~(\ref{d-k}). For $k \le m_g$  (with $\mu = m_g$) one gets 
\begin{equation}
 f(k) = {{f(m_g)} \over  {\left[ 1 + {{5\bar\beta_0} \over {3(4\pi)^2}} 
 d^2(m_g) \left( {{ k - m_g} \over {m_g}} \right)\right]^{1\over 2} } } 
 = f(m_g) {{d(k)}\over {d(m_g)}},
\label{fles}
\end{equation}
and for $k \ge m_g$, 
\begin{equation}
 f(k) = {{f(m_g)} \over  
 { \left[1 + {{\bar\beta_0} \over {(4\pi)^2}} d^2(m_g) \log({k^2\over m_g^2})  
   \right]^{1\over 2} } } = f(m_g)
 {{d(k)}\over {d(m_g)}}.
\label{fgt}
\end{equation}
The freedom in choosing the normalization for $Z_K(\Lambda)$
is now reflected in the unspecified normalization constant $f(m_g)$. 
The maximal infrared enhancement
of the Coulomb kernel is given by $k^{-7/2}$ ($f(k) \propto
1/k^{1/2}$)  if the
approximate solution of Eq.~(\ref{klg}) is used, or 
is given by $k^{-15/4}$ ($f(k) \propto 1/k^{3/4}$) 
if the full solution for d(k), Eq.~(\ref{klgex})
is used. We note that a linearly rising Coulomb potential requires
$f(k) \propto 1/k$ for small $k$. The exact numerical behavior of
$f$ will be discussed in the next subsection.
Lastly, if one substitutes the ansatz solution for the gap function
Eq.~(\ref{omaprox}) into  the gap equation (\ref{renom}), one finds that
it is indeed a solution up to terms of order ${\cal O}(k/m_g)$
for $k<m_g$ or ${\cal O}(m_g/k)$
for $k>m_g$.

To summarize, the approximate analytical solution leads
to a running coupling (FP operator), $d(k)$ which
falls off logarithmically at large momenta and  is enhanced at
small momenta. The approximate solution indicates that there is
only one critical value of the coupling for which the enhancement is
maximal and given by $d(k) \propto 1/\sqrt{k}$. This may be an 
artifact of the truncation of the series of coupled self-consistent
equations. One expects; however,  that the critical behavior
is universal, {\it i.e.} near the critical coupling higher order
corrections to the vertices in the Coulomb operator become irrelevant. 

The full Coulomb kernel becomes logarithmically suppressed at
large momenta as
expected from an all-order resummation of leading logs. At the
critical point and for low momenta it becomes enhanced over the perturbative
$1/k^2$ behavior and scales as $k^{-15/4}$. 
We have thus obtained a tantalizing glimpse of the possibility of constructing
a phenomenologically viable truncation of QCD.

\subsection{Numerical Solution} 

Encouraged by the near-appearance of linear confinement in the approximate 
analytical solution we proceed to a full numerical solution to the
truncated renormalized coupled gap equations. The solution is obtained
by mapping the gap equations onto a set of discrete nonlinear equations
by placing all functions on a momentum space grid. We have
found that numerical stability is enhanced if the grid is chosen
carefully, in particular by preferentially populating the low and high 
momenta regions.
The discrete gap equations were then solved with two independent 
 solution algorithms. Both methods used an iterative procedure
to cycle through the three equations. Convergence was typically achieved
in only a few passes since the analytical starting point
of the last section is quite accurate.

The numerical and approximate analytical solutions for the FP operator
are shown in Fig. 4 for three separate values of $d(m_g)$. This and
subsequent figures are plotted in units of $m_g$ which 
after renormalization is the only dimensionful parameter. Its value can 
 only be determined
upon comparison to a physical observable.
It is clear that the analytical solutions are
very accurate. Furthermore, the existence of  a critical coupling appears
to be numerically confirmed, with a value very near $d_c = 3.5$. 

\begin{figure}[hbp]
\hbox to \hsize{\hss\psfig{figure=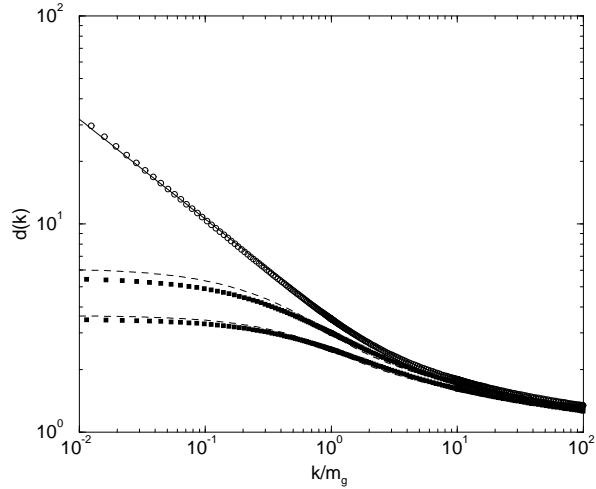,width=3.5in,angle=0}\hss}
\vspace{0.5cm}
\caption{Solution for the expectation value of the FP operator $d(k)$.  
The two lower dashed lines correspond to analytical, approximate 
solution with  $d(m_g)=2.5$ (lower)  and
$d(m_g)=3$ (higher). Boxes correspond to a full
numerical solution for the corresponding values of $d(m_g)$. The
numerical solution close to the critical point is shown by the open
circles. The solid line corresponds to a fit to this numerical
solution using the formula in Eq.~(\ref{fitd}) }
\label{Fig3}
\end{figure}

The numerical solution near the critical point has been fit to the
formula
\begin{equation}
d(k) = \left\{ \begin{array}{cc} d_c \left({m_g \over k}\right)^{a_d} & \mbox{ for } k
    \le m_g ,\\
     d_c \left( {{ \log( 1 + b_d ) } \over {\log(k^2/m_g^2 + b_d) }} 
\right)^{c_d} & \mbox{ for } k \ge m_g. \end{array} \right. \label{fitd}
\end{equation}
The fit yields $d_c = 3.5$, $a_d = 0.48$, $b_d= 1.41$ and $c_d =
0.4\;$ verifying the accuracy of the approximate analytical solution. 
Fig. 5 shows the Coulomb kernel function $f(k)/f(m_g)$ 
for $d(m_g) = 2.5$ and $3.0$.
Again, for $d(m_g) < d_c$ the solution saturates 
at low momentum and the analytical
approximation is quite accurate. The solution at the critical point is
compared with 

\begin{equation}
f(k)/f(m_g) = \Bigg\{ \begin{array}{cc} \left({m_g \over k}\right)^{a_f} & \mbox{ for } k
    \le m_g, \\
      \left( {{ \log( 1 + b_f ) } \over {\log(k^2/m_g^2 + b_f) }} 
\right)^{c_f} & \mbox{ for } k \ge m_g .\end{array}  \label{fitf}
\end{equation}
The fit yields, $a_f = 0.97$, $b_f=0.82$ and $c_f=0.62$.
The low momentum behavior is found to be more enhanced than in the
approximate analytical solution. The two fits to the numerical solutions
for $d$ and $f$ result in the following expression for the Coulomb
kernel $V(k) = f(k)d^2(k)/k^2$:

\begin{equation}
k^2 V(k)/f(m_g) = \left\{ \begin{array}{cc} (3.50)^2
  \left({m_g \over k}\right)^{1.93} 
 & \mbox{ for } k
    \le m_g ,\\
      8.07 \log^{-0.80}(k^2/m_g^2 + 1.41)
   \log^{-0.62}(k^2/m_g^2 + 0.82) 
  & \mbox{ for } k \ge m_g \end{array} \right. . 
\label{fitv}
\end{equation}

\begin{figure}[hbp]
\hbox to \hsize{\hss\psfig{figure=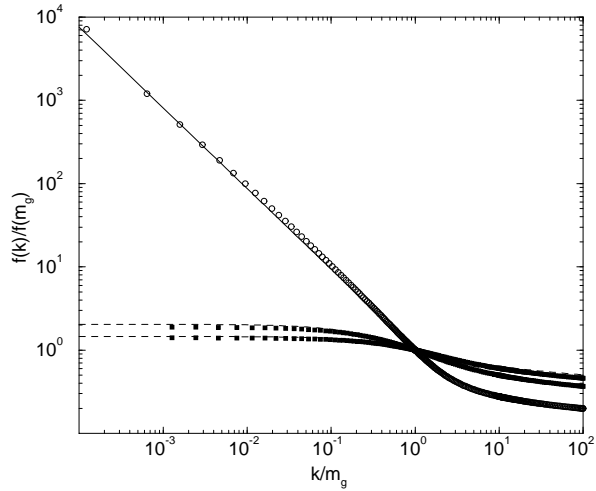,width=3.5in,angle=0}\hss}
\vspace{0.5cm}
\caption{$f(k)/f(m_g)$. Curves as in Fig. 4. The numerical
  solution near the critical point (open circles) is fit to the formula
  given by Eq.~(\ref{fitf}) (solid line). }
\label{Fig4}
\end{figure}

At low momenta the effective coupling $\alpha_{\rm eff}(k)$  (defined through 
Eq.~(\ref{alphaeff})) behaves very nearly as 
$1/k^2$. The fact that the power is not exactly $-2$ may be due to discretization
error (a finer momentum grid does indeed bring the coefficient closer to $-2$)
or the truncations employed in deriving the gap equations. In any
event, as will be shown later 
  the difference (roughly 3.5\%) is completely negligible with regards
to phenomenology.

Assuming linear confinement ($2a_d + a_f = 2$) 
 gives~\cite{Szczepaniak:2000uf}
\begin{equation}
6\pi b = (3.5)^2 m_g^2.
\end{equation}
Inserting the quark model value for the string tension, $b = 0.18$ GeV$^2$
yields $m_g \approx 530$ MeV.  Alternatively, lattice string tensions are
typically $0.26$ GeV$^2$\cite{JKM}, giving 
$m_g \approx 630\mbox{ MeV}$.  These estimates of the scale are in accord
with lattice computations of the adiabatic hybrid surfaces (thus is discussed
further in Sec.~IV.A) and with old glueball phenomenology\cite{gbs}.

 The numerical and ansatz solutions for the gap function are
shown in Fig.~6. We note the remarkable accuracy of the simple ansatz
for $\omega$, the main difference being the smooth transition
through the intermediate momentum region. Notice also that $\omega$
approaches $k$ very rapidly for large momentum.

\begin{figure}[hbp]
\hbox to \hsize{\hss\psfig{figure=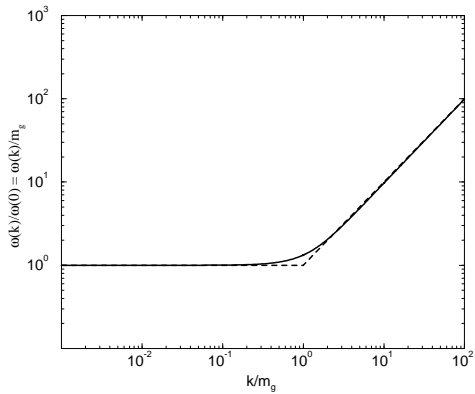,width=2.8in,angle=0}\hss}
\vspace{0.5cm}
\caption{ Comparison of the analytical approximation to $\omega(k)$
  (dashed line) and the full numerical solution (solid line).} 
\label{Fig6}
\end{figure}

Finally, the numerical stability of the solutions have been tested by
varying the number of grid points. Of course this also tests the {\it de facto}
numerical cutoff dependence of the results.
The results are shown in  Fig.~7. We find that the numerical results are
stable to within a  percent. Notice that this also confirms that
 all UV divergences have been properly subtracted.

\begin{figure}[hbp]
\hbox to \hsize{\hss\psfig{figure=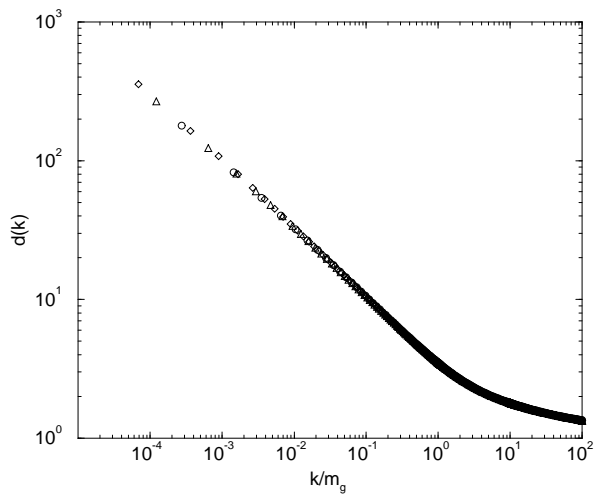,width=3.5in,angle=0}\hss}
\vspace{0.5cm}
\caption{Numerical solution for $d(k)$ near the critical point for
  $192$ (circles), $288$ (triangles) and $384$ (diamonds) grid points.}
\label{Fig7}
\end{figure}




\subsection{Higher Order Terms}

We now address the issue of the neglected terms in the coupled gap equations.
These  arise, for example,  from truncation of the rainbow-ladder sums, 
 higher order corrections to the Coulomb vacuum
 energy, and from the terms generated by the
 Faddeev-Popov determinant ${\cal J}$.

\subsubsection{Vertex Corrections}

The truncation to the rainbow-ladder resummation for the Faddeev-Popov
operator and the Coulomb kernel ignores higher order, ${\cal O}(d^n)$, $n\ge
2$ corrections to the triple Coulomb-transverse-gluon vertex. Using
the approximate analytical solutions for $d(k)$ and $\omega(k)$ we 
estimate these contributions by evaluating the ${\cal O}(d^2)$ correction. From
Eq.~(\ref{gamma1}) it follows that the lowest order correction to the
bare vertex is given by 

\begin{equation}
 \delta \Gamma^{c}_i(\k,\q,\k-\q) =  
  {N_c\over 2} 
   \int {{d\l}\over 
{(2\pi)^3}} 
 {{ [(\k + \l) \cdot \delta_T(\q)]^i[\k\cdot \delta_T(\l) (\k + \l
  -\q)] }
\over {2\omega(\l)}}
 {{d(\k+\l)}\over {(\k+\l)^2}} 
 {{d(\k+\l-\q)}\over {
\k+\l-\q)^2}} T^c.
 \label{dgamma1}
\end{equation}
We have evaluated this integral numerically and found that for all
values of the external momenta the correction does not exceed a few
percent.

\subsubsection{Second and Fourth Order Corrections to the Coulomb Kernel}

Recall that an operator product expansion for the Coulomb kernel has been
defined in Sec.~II.E, Eq.~(\ref{Kfull}). We now employ the Swift equation
(\ref{swift}) and the operator expansion of the Faddeev-Popov operator 
Eq.~(\ref{dfull}) to derive an explicit expression for the terms in that
expansion:

\begin{eqnarray}
K^{ab}(\k,\p;\A) &=& \delta(\k+\p) \delta^{ab} K^{(0)}(\k) + i g f^{acb} \left[ 
K^{(0)}(\p) D^{(0)}(\k) + D^{(0)}(\p)K^{(0)}(\k) \right] : \A^c(\p+\k) \cdot \p:  + \nonumber \\
&+&  \ldots + (i)^n f^{a c_1 e_1} \ldots f^{e_{n-1}c_n b} {d\over dg}\left[
g^{n+1} D^{(0)}(\p) D^{(0)}(\p-\s_1) \cdots D^{(0)}(\p - \sum_{\ell=1}^n \s_\ell) \right] \nonumber \\
 && \delta(\p +\k - \sum_{\ell=1}^n \s_\ell)
 : \A^{c_1}(\s_1) \cdot \p \ldots \A^{c_n}(\s_n) \cdot (\p - \sum_{\ell=1}^{n-1} \s_\ell) :
\label{kexp}
\end{eqnarray}

\noindent
The term in the expansion of $K$ which contains $n$ gluons
is weighted by a product of $n-1$ factors of $D^{(0)}$ and a single
factor of $K^{(0)}$.
The additional contributions to the VEV of the Hamiltonian discussed
in Sec.~II.E, ${\cal E}^{(2)}_C$ and  ${\cal E}^{(4)}_C$, come from 
terms with a product of  $n=2$ and $n=4$ normal ordered gluon fields 
$:\A^n:$ respectively.  These are the only contributions which have a 
nonzero VEV 
after combining with the charge densities.
The contribution to the gap equation is
then obtained by taking the derivative of the VEV 
with respect to $\omega$. As was discussed earlier, an alternative 
method to derive the gap
equation is to require that the off-diagonal (proportional
to $\alpha^{\dag}\alpha^{\dag}$ or $\alpha\alpha$) portions of the 
one-body operators vanish. The second  method would indicate 
that terms with $n=6$, $\rho:\A^6:\rho$,
contribute to the gap equation as well since the four gluon fields from the two charge
densities can contract with the fields from the kernel leading to an
operator proportional to $:\A^2:$. As discussed in Sec.~II.E, the 
apparent difference in these two procedures is resolved if one notices 
that there are contributions to the gap equation which arise from 
the implicit dependence of the Coulomb kernel on $\omega$. In the second
method, the
contribution which would be associated with the
$n=6$ term in the operator product expansion of $K$ is identical to
the one from  the 
derivative of the kernel
in the $n=4$ term contribution to the VEV. This was denoted 
${\cal E}^{(4),K}_C$ in Sec.~II.E.
Similarly the term
referred to as ${\cal E}^{(0),K}_C$ in the discussion preceding 
Eq.~(\ref{gapgen}) 
is identical to the contribution
from the $n=2$ term when the fields from the charge
densities are contracted with each other. 

Adding all these pieces together yields, 

 \begin{eqnarray} 
 \omega^2(q) - \omega(\mu)^2  = & &  q^2  
  + [{\cal E}^{(0),\omega}_C(q)]
   + [{\cal E}^{(0),K}_C(q) + {\cal E}^{(2),\omega}_C(q)]  
   + [{\cal E}^{(2),K}_C(q) + {\cal E}^{(4),\omega}_C(q)] \nonumber \\
 & &   + [{\cal E}^{(4),K}_C(q)] - (q \to \mu).  
\label{fullgapeqn}
\end{eqnarray}
The four terms in the brackets are  ${\cal O}(d^2)$, ${\cal O}(d^4)$, 
 ${\cal O}(d^6)$ and ${\cal O}(d^8)$ respectively; no other corrections exist.
We test the importance of the higher order terms by computing the 
${\cal O}(d^4)$ correction to the truncated gap equation.

\begin{figure}[hbp]
\hbox to \hsize{\hss\psfig{figure=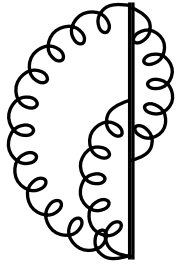,width=1.5in,angle=0}\hss}
\vspace{0.5cm}
\caption{ ${\cal O}(d^4)$ contributions to the gap function 
 from ${\cal E}^{(2),\omega}_C$.}
\label{Fig8}
\end{figure}

An example of a diagram contributing to ${\cal E}^{(2),\omega}_C$ is given in 
Fig.~\ref{Fig8}. The explicit expressions for ${\cal E}^{(0),K}_C$ and
${\cal E}^{(2),\omega}_C$ are given below.

\begin{eqnarray}
{\cal E}^{(0),K}_C(q) & = &   
   { {N_c^2}\over {8}} 
          \int  \dk \ddp 
           \left[ {{\omega(k)}\over {\omega(p)}}-1\right]
           Tr [\delta_T(\hat\p)\delta_T(\hat\k)] 
           [(\k + \p)\delta_T(\hat\q)(\k+\p)]
            \nonumber \\
& & \times     
           \left[ {{f(\k+\p)d^2(\k+\p)d(\k+\p)d(\k+\p+\q) + perm.} 
            \over { (\k+\p)^2(\k+\p)^2(\k+\p+\q)^2 } } \right] 
            \nonumber \\ 
\end{eqnarray}
and 
\begin{eqnarray}
{\cal E}^{(2),\omega}_C(q) & = &   { {N_c^2}\over {16}} 
          \int  \dk \ddp \left[1 - {{\omega^2(q) }\over
          {\omega(k)\omega(p)}} 
  \right]
           [(\q + \k)\delta_T(\hat\p)
           \delta_T(\hat\q)\delta_T(\hat\k) (\q+\p)] 
            \nonumber \\
        & & \times     
           \left[ {{f(\q+\k+\p)d^2(\q+\k+\p)d(\q+\k)d(\q+\p) + perm.} 
            \over { (\q+\k+\p)^2(\q+\k)^2(\q+\p)^2 } } \right]
          \nonumber \\
 & & + { {N_c^2}\over {8}} 
          \int \dk \ddp 
           \left[1- {{\omega(k)}\over {\omega(p)}}\right]
            (\k-\q)\delta_T(\p)\delta_T(\k)\delta_T(\q)(\p-\k)
            \nonumber \\
        & & \times     
           \left[ {{f(\p-\k)d^2(\p-\k)d(\p-\k+\q)d(\k-\q) + perm.} 
\over { (\p-\k)^2(\p-\k+\q)^2(\k-\q)^2 } } \right], \nonumber \\     
\end{eqnarray} 
Here the permutations refer to the other two ways of arranging the
argument of $fd^2$ in $f(1)d^2(1) \cdot d(2) \cdot d(3)$.

\begin{figure}[hbp]
\hbox to \hsize{\hss\psfig{figure=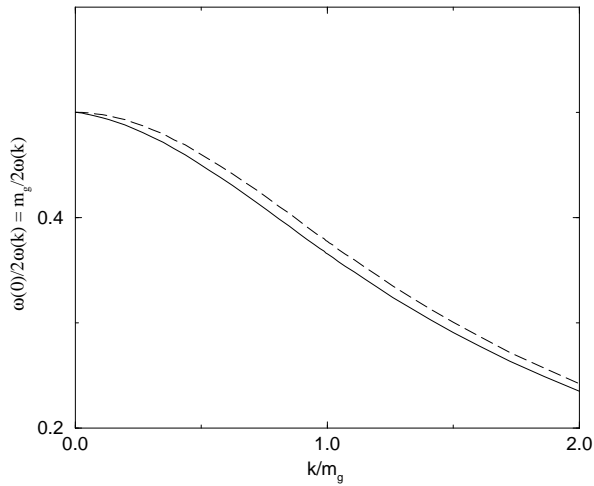,width=3.5in,angle=0}\hss}
\vspace{0.5cm}
\caption{Normalized  instantaneous transverse gluon propagator,
  $1/2\omega(k)$. The dashed line is the solution to the leading order gap
equation of Eq.~(\ref{renom}); the solid line includes  ${\cal O}(d^4)$ 
corrections.}
\label{newgap}
\end{figure}

Including these terms in the gap equations modifies the results for $d$
and $f$ by strengthening the IR enhancement somewhat.
The result for the 
gap function is shown in Fig. \ref{newgap}. As expected, the change at higher
momenta is minimal. However, we see that the ${\cal O}(d^4)$
terms do not modify $\omega$ at low momentum either. This is because
  ${\cal E}^{2,\omega}$ and ${\cal E}^{0,K}$ depend on the combination 
  $\omega(p_1)/\omega(p_2)-1$ which suppresses them in the IR limit ($\p_1 = \p_2$).
Our results are compared to lattice computations
 in Sec.~IV.A. We stress
that the result of Fig.~9 should be considered preliminary because
 there are other ${\cal O}(d^4)$ corrections (see the next
 subsection)  that have not yet been included. 

The computation of  the ${\cal O}(d^6)$ and ${\cal O}(d^8)$
corrections
is progressively more difficult
and is currently under investigation. These require the
numerical solution of a self-consistent
equation involving at least 8-dimensional integrals. However, since the
${\cal O}(d^4)$ corrections are small 
we expect these higher order
terms not to change the results significantly.

\subsubsection{Faddeev-Popov Contributions}

We now discuss the corrections due to the Faddeev-Popov determinants 
${\cal J}$. We calculate the contribution to the gap equation
for the determinant present in the 
kinetic part of the Hamiltonian, through the 
${\cal J}^{-1/2}\Pi {\cal J}^{1/2}$ operators. This is given by 

\begin{equation}
{1\over 2} \int d\x {\cal J}^{-1/2} \Pi^a(\x) {\cal J} \Pi^a(\x) {\cal
    J}^{1/2} = {1\over 2} \int d\x \Pi^a(\x) \Pi^a(\x) + V_A.
\end{equation}
Similarly, $V_B$ is defined via the relation (see Eq.~\ref{hc2})

\begin{equation}
H_C = {1\over 2}\int d\x d\y \rho^a(x) K_{ab}(\x,\y;\A) \rho^b(y)  + V_B.
\end{equation}

A direct computation yields 
\begin{eqnarray}
V_A = & &  {g^2\over 4}\int d\x  f^{abc} f^{aef} \delta_T(\bdel_\x)_{ij} 
\langle \x b|(\bdel \cdot D)^{-1} |\x, c\rangle
\stackrel{\leftarrow}{\bdel_\x}_j
 \delta_T(\bdel_\x)_{ik} 
\langle \x e|(\bdel \cdot D)^{-1} |\x, f\rangle
\stackrel{\leftarrow}{\bdel_\x}_k \nonumber \\
& - &  {g^2\over 8}\int d\x  f^{abc} f^{aef} \delta_T(\bdel_\x)_{ij} 
\langle \x b|(\bdel \cdot D)^{-1} |\x, f\rangle
\stackrel{\leftarrow}{\bdel_\x}_j
 \delta_T(\bdel_\x)_{ik} 
\langle \x e|(\bdel \cdot D)^{-1} |\x, c\rangle
\stackrel{\leftarrow}{\bdel_\x}_k. \nonumber \\
\end{eqnarray}
or in momentum space 
\begin{eqnarray}
V_A = &&-{1\over 8} \int {d^3 k\over (2 \pi)^3} {d^3 p \over (2\pi)^3}
{d^3 q \over (2\pi)^3}
D^{ac}(\q,\q+\k) D^{de}(\p+\k,\p) f^{bca}f^{bed} [\q
\delta_T(\k)\p] \nonumber \\
&& + {1\over 4}  \int {d^3 k\over (2 \pi)^3} {d^3 p \over (2\pi)^3} {d^3
q \over (2\pi)^3}
D^{ac}(\q,\k+\p) D^{de}(\p,\q-\k) f^{bcd} f^{bea} [\q \delta_T(\k)
\p] \\
\end{eqnarray}
where $D^{ac}(p,k) = \langle a \p | {g \over \nabla \cdot D} | c
\k\rangle$. 
We note that $V_A$ is similar to Christ and Lee's $V_1$; however, it is
not identical because we have not Weyl ordered the Hamiltonian.

Using the operator product expansion for the FP operator, 
 these lead to terms proportional
 to $:\A^2:$ which add to the gap equation the following
contribution 

\begin{eqnarray} 
{\cal E}^{2,\omega}_{FP}(q) = & &  { {N^2_C}\over 16}\int \dk \ddp
 {{ d(\k+\q)}\over {(\k+\q)^2}} 
 {{ d(\k)} \over {\k^2}} [\k \delta_T(\q) \p]^2 
 {{ d(\p+\q)} \over {(\p+\q)^2}}
 {{ d(\p)} \over {\p^2}} \nonumber \\ 
 &  - & { {N^2_C}\over 16} \int \dk \ddp 
 {{ d(\k+\q)}\over {(\k+\q)^2}} 
 {{ d(\k)} \over {\k^2}} [\k \delta_T(\q) \p] 
[(\k+\q)\delta_T(\p+\q+\k)(\p+\q)]
 {{ d(\p+\q)} \over {(\p+\q)^2}}
 {{ d(\p)} \over {\p^2}}  \nonumber \\
 &  - & { {N^2_C}\over 4} \int \dk \ddp 
 \left({{ d(\k)} \over {\k^2}}\right)^2 [\k \delta_T(\p) \k] [\k
 \delta_T(\q) \k ]  
 {{ d(\k+\q)} \over {(\k+\q)^2}}
 {{ d(\k+\p)} \over {(\k+\p)^2}} 
\end{eqnarray}

The contribution of  $V_A$ to the gap equation is 
IR-finite but UV-divergent thus will modify the gluon mass
counterterm. 
A detailed numerical study of
the full $O(d^4)$ corrections to the gap equation will be presented
elsewhere.

\section{Discussion}

As demonstrated in the previous section, the asymptotic behavior of the
numerical solution to the gap equations is  $V(k) \sim 1/k^4$, it thus
appears that the methodology advocated in this paper is capable of
describing quark confinement. The appearance of the confinement
phenomenon hinges crucially on the choice of the variational vacuum
which we use to construct the quasiparticle basis and on realizing that
this choice also affects the interaction between these quasiparticles
via the summed expression for the instantaneous  Coulomb kernel. We now
examine the implications of this success on
confinement and the Gribov ambiguity.

\subsection{The Confinement Potential}

The requirement that the gluon mass gap function $\omega(k)$ be cutoff
independent gives rise to a mass scale which we call the gluon mass, $m_g$.
The value of $\omega$ at a particular momentum scale, say $k=0$ serves 
then as the
underlying mass parameter of the theory. At the critical coupling $d_c$
the only free parameters in the gluon sector are $Z_K(\mu)$
and the momentum scale itself, $m_g=\omega(0)$. Nonperturbative
renormalization may be carried out by requiring that the
Coulomb kernel reproduce the static $Q\bar Q$ heavy quark
potential as seen on the lattice (recall that $V_{Q\bar Q}$ is
a renormalization group invariant quantity). In our approach this
potential is given by 

\begin{equation}
H_{QCD} |Q(r/2),{\bar Q}(-r/2) \rangle = V_{Q\bar Q}(r)|Q(r/2),{\bar Q}(-r/2) \label{vqq}
\rangle,
\end{equation}
In pure QCD, {\it i.e.} ignoring light flavors, the above  eigenstate can be 
expanded in terms of multigluon states constructed from the
quasiparticle operators 
acting on the $|\omega\rangle$ vacuum. Schematically,

\begin{equation}
|Q(r/2),{\bar Q}(-r/2) \rangle = \sum_{n=0} \psi_n
  (\alpha^\dagger)^n b^{\dag} (r/2) d^{\dag}(-r/2)  |\omega\rangle .
\end{equation}
where the quark creation operators refer to static sources. 
The Hamiltonian mixes states differing by gluon number; however,  
one expects that the mixing 
between such states to be suppressed by energy denominators 
due to the gluon mass gap, $\omega(0) \ne 0$
(this is 
discussed in much more detail in Sec.~V.A).
This mass gap can be estimated from the difference between the lowest
and excited  adiabatic $Q\bar Q$ potentials which have been 
calculated on the lattice\cite{JKM}. 
One finds that this difference is
 $\Delta V(r\sim 1\ \mbox{fm})  \sim 600 - 800\ \mbox{MeV}$. This is
a natural estimate for $\omega(k)$ at low momenta. 
The implication is that the static ground state heavy quark potential 
may be accurately
computed by ignoring extra gluonic excitations in the heavy quark system.
(A calculation of the 
excited adiabatic potentials will be presented elsewhere.)
Thus, to good accuracy, the static heavy quark potential $V(r)$ is given by

\begin{equation}
V_{Q\bar Q}(r) = -C_F V(r) = -C_F \int \dk e^{i\k\cdot \x} {{f(k) d^2(k)} 
\over {\k^2}} \equiv -C_F \int \dk e^{i\k\cdot \x} 
{{4\pi \alpha_{\rm eff}(k)} \over {\k^2}}.
\end{equation}

It is useful to return to the approximate analytical solutions to
the truncated renormalized gap equations of Sec. III.D to illustrate 
how the different parameters enter. 
We have seen that at the critical point $d_c$ the solutions for the
Faddeev-Popov operator and
the Coulomb kernel are 

\begin{equation}
 d(k) = \left\{ \begin{array}{cc} 
  d(m_g) \left( { {m_g} \over k }\right)^{1\over 2}  & \mbox{ for } k
  \le m_g \\
    {{d(m_g)} \over {\sqrt{ 1 + {3\over 5} \log {{k^2}\over {m_g}^2} }}}
    & \mbox{ for } k \ge m_g  \end{array} \right. \label{ddass}
\end{equation} 
 and 
\begin{equation}
 f(k) = \left\{ \begin{array}{cc} 
 f(m_g) \left( { {m_g}  \over k }\right)^{3\over 4} & \mbox{ for } k \le m_g \\
   {{f(m_g)} \over {\sqrt{ 1 + {3\over 5} \log {{k^2}\over {m_g}^2} }}}
    & \mbox{ for } k \ge m_g  \end{array} \right. \label{fass} 
\end{equation}
Since $d(m_g) = d_c \sim 3.4414$ is fixed, the potential has only two free
parameters, the overall strength determined by 
$f(m_g)$ and the mass scale set by $m_g=\omega(0)$. These
may be determined by comparing with lattice computations of the Wilson loop.
One finds $f(m_g) \sim 1.0$ and $m_g \approx 1.8/r_0$. Here $r_0$ is the
Sommer parameter of lattice gauge theory which is determined to be
roughly 1/430 MeV$^{-1}$. Thus $m_g \approx 720\mbox{ MeV}$.

\begin{figure}[hbp]
\hbox to \hsize{\hss\psfig{figure=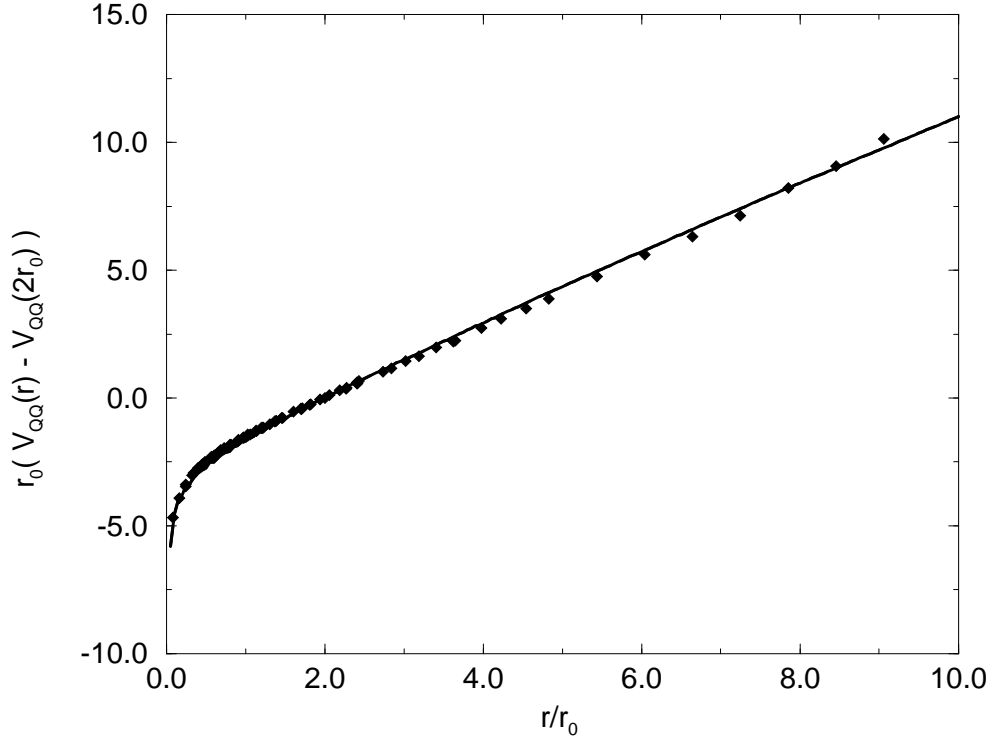,width=5.5in}\hss}
\vspace{0.5cm}
\caption{ Static $Q{\bar Q}$ ground state potential. 
 The solid line is the full numerical solution for $f(m_g)=1.41$ and 
 $m_g = 1.4/r_0 = 600\mbox{ MeV}$. 
 Data are taken from Ref.~[22].}
\label{Fig10}
\end{figure}

The same procedure may be followed for the numerical solution to the
gap equation. Good agreement with
the lattice static potential is obtained by choosing $f(m_g) = 1.41$
and $m_g = 1.4/r_0 = 600$ MeV. The minimum in parameter space is fairly
broad, for example $f(m_g) = 1.09$ and $m_g = 1.6/r_0 = 690$ MeV 
provides nearly as good a description of $V_{Q\bar Q}$. 
The resulting potential (after numerically Fourier
transforming to configuration space) is presented in Fig.~\ref{Fig10}. One
sees that the numerically obtained static quark potential 
provides a reasonable facsimile of the lattice potential. 
This somewhat surprising result provides {\it a posteriori} support
for the methodology advocated here.

\subsection{The Gribov-Zwanziger Horizon and the Gap Function}
 
As mentioned in Sec.~II.B, 
the Gribov problem may be resolved by selecting a single gauge copy 
from the ensemble of Gribov copies by imposing the horizon
condition of Eq. (\ref{horizon}). Furthermore, Zwanziger \cite{Cucchieri:1997ja} has 
shown that the restriction to the fundamental modular region imposed
by the horizon term is equivalent to having a low-momentum enhancement in
the Faddeev-Popov operator over the perturbative $1/k^2$
behavior. This enhancement takes the form
\begin{equation}
{ 1 \over {D^{(0)}(k)}}  
= k_i k_j( \Sigma(0)_{ij} - \Sigma(k)_{ij}),
\label{FMR}
\end{equation}
where $\Sigma$ is a regular function at the origin.
Such a behavior is clearly an indication of confinement~\cite{gribov}, 
since the FP operator determines the static potential
between color sources (cf. Eq.~(\ref{hc2}) or Eq.~(\ref{swift})).
Comparing  Eq.~(\ref{FMR}) with Eq.~(\ref{ddef}) shows that this is
equivalent to the statement that $d(k)$ is singular at the origin,

\begin{equation}
\lim_{k \to 0} 1/d(k) \to 0.
\end{equation}
The behavior of $d$ at small momenta 
depends on one integration constant, $d(\mu)$. As we have shown 
earlier, $d(k)$ approaches a finite value as $k\to 0$ except when
$d(\mu) = d_c$ where the Gribov-Zwanziger singularity develops.

It is possible that the saturation of $d(k)$ to finite values when $d(\mu) < d_c$
is an artefact of the rainbow-ladder approximation -- we leave this as a
matter of future investigation. For the present we simply require that the
theory give rise to an enhancement of the FP operator at small momentum,
this boundary condition then selects the coupling $d(\mu) = d_c$.

In Ref.~\cite{Cucchieri:1997ja} the enhancement in the FP
operator was obtained by adding the horizon term to the Hamiltonian
via a Lagrange multiplier. The VEV of the new Hamiltonian was then
computed in the bare vacuum, {\it i.e} with $\omega(k) = k$; however,
the horizon terms adds a mass term through an effective $\A^2$ operator 
whose strength
is determined by the expectation value of the FP operator. This term
has the effect of enhancing $d(k)$ for small $k$. The
equivalent of Eq.~(\ref{dgen}) was then solved and the Coulomb operator was 
approximated by the square of the FP operator.

In our approach the horizon condition was used to justify the 
expansion of the FP operator in a power series in $g\A$. The resulting
expressions for the FP operator were summed in the presence of a nontrivial
mean field background (the variational vacuum) with the aid of the rainbow
approximation producing IR enhanced FP and Coulomb VEVs. The success of this
procedure demonstrates that the explicit effects of the horizon term may
be ignored if one is willing to develop the quasiparticle spectrum and 
interaction self consistently.






To further test this mechanism for realizing the Gribov-Zwanziger confinement
scenario we compare our result for the gap function to that computed by
Cucchieri and Zwanziger in SU(2) lattice gauge theory\cite{CZ}. In that paper, the authors measure the transverse and 
instantaneous gluon propagators in `minimal Coulomb gauge'. They compare
the numerical results to a functional form proposed by Gribov\cite{gribov}:

\begin{equation}
D^{tr}(k) = {1\over {2E_G(k)}}, \; E_G(k) = {1\over k} \sqrt{k^4 + M_G^4}.
\end{equation}
We call the scale appearing in this relationship, the Gribov mass, $M_G$.
Cucchieri and Zwanziger found the the computed instantaneous transverse
propagator agreed very well with this functional form but does not
reproduce the normalization. 

As discussed earlier, in our approach the transverse gluon propagator 
is suppressed
due to an infrared singularity in the one-body gluon
operator. Explicitly, the one body operator in the quasiparticle basis
is given by 

\begin{equation} 
H_{one-body}  = \sum_{\lambda,c} \int \dk E(k) \alpha^{\dag}(\k,\lambda,c)
\alpha(\k,\lambda,c) 
\end{equation}
with 
\begin{equation} 
E(k) = \omega(k) \left[ 1 + {N_c\over 4}\int \dq
  {{f(\k-\q)d^2(\k-\q)}\over {(\k-\q)^2}} {{ 1 + ({\hat \k}\cdot{\hat
  \q})^2}
 \over {\omega(q)}} \right].
\end{equation}
The low-momentum enhancement of the kernel makes the integral
infrared singular. Thus, as expected, gluons do not propagate. 

We note that the equal time transverse gluon
propagator is not determined by $E(k)$ but by $\omega(k)$

\begin{equation}
D^{tr}(x) = 
 \lim_{t \to 0} \langle \omega|  T[ \A^a(\x,t) \A^b({\bf 0},0)] |\omega
\rangle 
 = \delta^{ab} \int \dk { {\delta_T(\k)} \over {2\omega(k)}} e^{i\k\cdot \x} .
\end{equation}

We have seen that the gap function obtained in Sec.~III is rather flat
 at small momenta, even when 
some $d^4$ corrections are incorporated into the gap equation. This is
 inconsistent with the lattice calculation for the same quantity, but
 as shown above, 
 not inconsistent with the Gribov confinement scenario. The
 disagreement with lattice may
 be  due to the use of the rainbow-lattice approximation and is being
 investigated.  It is worth noting however that the effective
  gluon mass found by comparison to the $Q{\bar Q}$ potential (or
 alternatively the $D^{00}$ gluon propagator) is consistent with that
 found in the Coulomb gauge lattice calculations. 

\section{Implications for the Constituent Quark Model and Phenomenology}

We now turn to an examination of the implications of the results
presented here on 
the phenomenology of hadrons. Since this depends crucially on the
explicit definition of hadronic states, we begin by searching
for an efficient way to construct hadrons by specifying a new
constituent quark model of QCD.
The phenomenology of confinement is then analysed in light
of the results of the last two sections.  
We conclude with a clarification of several open issues
in the old constituent quark model and present a justification for the
surprising efficacy of the quark model for light hadrons.

\subsection{Constructing Hadrons}

It is clear that constructing hadrons from the basis of free quasiparticles
is futile if it is done perturbatively. A simple and natural way to avoid
this pitfall is to choose a convenient form of $H_0$ and diagonalise
it nonperturbatively to obtain a basis of color singlet bound states.
Bound state perturbation theory may then be employed to systematically
include the effects of $H_{int}$. In our case the natural assignment for
these operators is

\begin{equation}
H_0 = \int \psi^\dagger \left( - i {\bbox \alpha} \cdot \bdel + \beta m\right)
\psi + {1 \over 2}\int d\x  \bPi^2 - {1\over 2} \int d\x \A \bdel^2 \A
+ {1\over 2}\int d\x d\y\, \rho^a(\x) K^{(0)}(\x-\y) \rho^a(\y)
\label{h0}
\end{equation}
and
\begin{eqnarray}
H_{int} &= {1\over 2} \int d\x \left[ \B^2 + \A \bdel^2 \A \right]  -
g \int \psi^\dagger {\bbox \sigma}\cdot \A \psi + V_A + V_B +\nonumber \\
& +  {1\over 2}\int d\x d\y\, \rho^a(\x) \left[ K^{ab}(\x-\y;\A) - \delta^{ab}K^{(0)}(\x-\y)
\right] \rho^b(\y)
\label{hint}
\end{eqnarray}

The general philosophy is clear -- $H_0$ generates hadronic bound states;
$H_{int}$ incorporates the corrections to these states due to transverse
gluon exchange, three and four gluon interactions, and
higher order contributions from the FP determinant and 
 instantaneous confinement potential. It
is worth stressing that $H_0$ is still a field theory and hence is considerably
tougher to solve than old fashioned quantum mechanical quark models. But there
are substantial advantages to adopting this approach. Foremost is that
$H_0 + H_{int}$ {\it is} QCD. Furthermore, $H_0$ is relativistic and incorporates
gluonic degrees of freedom. Thus it is possible to examine glueballs, hybrids,
and other gluonic phenomena in a coherent fashion.  The utility of the rearrangement
made in Eq.~(\ref{hint}) lies in the use of the variational vacuum to construct a
phenomenologically viable basis of quasiparticles. This has the direct effect of
greatly improving the Fock space convergence of any observable. As we have seen, it
also automatically generates the correct static potential upon
which to construct hadrons.
We have previously mentioned that $H_0$ generates states which are infrared
divergent if they are not color singlets (hence these are removed from
the spectrum). Conversely, all color singlets are IR finite.  Thus the
basis generated by $H_0$ contains no spurious color nonsinglet states
which would have to be removed by laborious iteration of $H_{int}$ and, in
fact, is expected to provide a reasonably accurate starting point for
hadronic spectrum computations.
As a practical note, 
the physics of the variational vacuum may be accurately approximated by simply using
dressed quarks and gluons when constructing hadrons. The constituent masses are roughly
200 MeV and 600-800 MeV respectively.  Finally, the spectrum generated by $H_0$ is
spin averaged in the sense that it only incorporates spin
 effects from relativistic corrections to the Coulomb potential.  
  Full spin splittings come from $H_{int}$. 

\begin{figure}[hbp]
\hbox to \hsize{\hss\psfig{figure=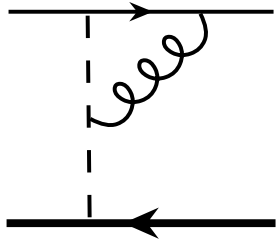,width=2.5in,angle=0}\hss}
\vspace{0.5cm}
\caption{The leading light quark correction to the confinement potential}
\label{qcorr}
\end{figure}

An important implication of this approach is that the rapid convergence
of the constituent quark model Fock space expansion has a natural  and
simple  explanation. All of the corrections induced by $H_{int}$ (for nonexotic
states)
involve the transfer of a virtual transverse gluon. Since these are
quasiparticles in the variational vacuum, the relevant perturbative
diagrams are suppressed by the mass gap between the regular
and hybrid states.
This simple feature of QCD in Coulomb gauge has important phenomenological
consequences.  For example, it implies that the Fock space expansion 
converges quickly because state mixing involves the creation of massive 
gluonic (or quark) quasiparticles.
Recently, lattice data has appeared which confirms this picture. 
Duncan {\it et al.}\cite{Duncan:1993eb} have constructed a simple
relativistic quark model of $B$ mesons by considering a light 
relativistic quark (with kinetic energy $\sqrt{k^2+m^2}$) moving in the
lattice $\Upsilon$ potential. Detailed comparison with lattice $B$ data 
demonstrated the high accuracy of the model. The point which is relevant
for our discussion is that the lattice $\Upsilon$ interaction (recall that 
this is equivalent in principle and in practice to $K^{(0)}$) should
receive corrections due to the light quark when applied to $B$ mesons; see
Fig.~\ref{qcorr}. The 
fact that these corrections are not important demonstrates that they
are suppressed, in agreement with the above arguments.

\subsection{Confinement in the Constituent Quark Model}

One of the benefits of Coulomb gauge is that it makes the source
of confinement clear: in the heavy quark limit quarks and transverse
gluons decouple  and the quark-antiquark quark interaction
arises solely from the instantaneous Coulomb operator. 
This rigorous result has several significant implications for hadronic
phenomenology.

\begin{figure}[hbp]
\hbox to \hsize{\hss\psfig{figure=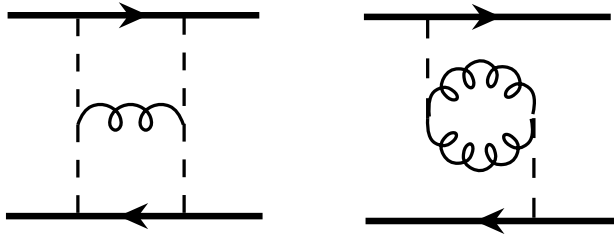,width=4in,angle=0}\hss}
\vspace{0.5cm}
\caption{Examples of Higher Order Corrections to the Heavy Quark-Antiquark
Interaction.}
\label{vcorr2}
\end{figure}

First we simplify the situation by noting that
higher order terms such as shown in Fig.~\ref{vcorr2} are suppressed due to the
arguments espoused in the previous subsection.
Thus, the dominate interaction between static color sources is the
leading kernel in the Coulomb interaction, $K^{(0)}$. As we have
seen, this kernel is essentially identical to the lattice Wilson
loop result, so this conclusion is supported {\it a posteriori}.

This simple statement carries wide repercussions. For example, a longstanding
cornerstone of quark model phenomenology is that confinement is `scalar'. What 
this means is that the interaction between quarks is assumed to be

\begin{equation}
{1\over 2} \int \bar \psi \psi(\x) K(\x-\y) \bar \psi \psi(\y).
\label{naive}
\end{equation}
This form (as opposed to `vector' confinement $\psi^\dagger \psi K \psi^\dagger \psi$) 
is supported by  
a comparison of the predicted spin splittings in heavy quarkonia with data\cite{schnitzer}.
However, the results presented here make it clear that this  conclusion is naive. The
interactions between color sources is more complicated than the simple
facsimile given in  Eq.~(\ref{naive}). As we have seen, the leading interaction between
quarks is given by $K^{(0)}$ -- and this has the form of vector confinement. What is taken
as evidence of the scalar nature of confinement is in fact quarkonium spin splittings 
which are
generated by nonperturbative mixing with intermediate hybrid states via $H_{int}$. That
this more complicated (and correct) picture may look `scalar' has been shown in 
Ref.~\cite{Szczepaniak:1997tk}

Another simple conclusion of the picture being developed here is that the 
confinement potential between color sources scales as the quadratic Casimir.
This follows from the observation that the dominant contribution to the 
confinement potential is given by the leading kernel and 
that the color structure of this kernel is 
  $K^{(0)}_{ab} = \delta_{ab} K^{(0)}$.
 The fact that Casimir scaling of the Wilson loop potential has been observed
 repeatedly\cite{casimir} may be taken as a successful prediction of our methodology or may be
used as further proof that the diagrams of Fig. \ref{vcorr2} are suppressed with
respect to $K^{(0)}$.

The methodology presented here allows for the resolution of several
open, but often ignored, ambiguities in the constituent quark model.
For example, it is often stated that the linear potential is built
from the exchange of infinitely many gluons. One may then ask
why the one gluon exchange potential is retained as an important
part of quark model phenomenology. Indeed the split between one gluon
exchange color Coulomb and hyperfine forces and the multiexchange
linear force is necessarily ambiguous. The resolution to this
issue is transparent in Coulomb gauge: `one gluon exchange' is
part of $H_{int}$ and is due to noninstantaneous transverse gluon exchange.
The instantaneous central portion of the quark model should consist of
a linear term in addition to the running resummed `Coulomb' term of
Eq.~(\ref{fitv}).  No ambiguity exists because of the separation of
instantaneous and transverse degrees of freedom inherent in Coulomb gauge.

Another problem with the old-fashioned CQM has to do with the previously mentioned
assumed scalar nature of confinement. Unfortunately,  scalar confinement 
implies that if mesons are bound by a linear potential,
baryons are antibound~\cite{Godfrey:1985xj}.
This is clearly
an intolerable situation  which is routinely ignored by CQM
practitioners. As we have seen, the resolution is that confinement acts
as the time component of a vector rather than as a scalar, and no
inconsistency exists between mesons and baryons.

\subsection{Constituent Gluons and Strong Decays}

We illustrate the power of our approach by considering
the vexing problem of strong decays in hadronic physics.

\begin{figure}[hbp]
\hbox to \hsize{\hss\psfig{figure=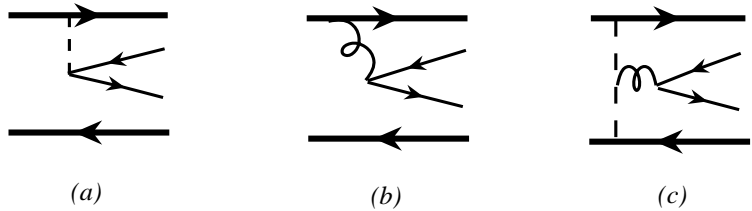,width=4.5in,angle=0}\hss}
\vspace{0.5cm}
\caption{Leading Order in $1/m_g$ Meson Decay Diagrams}
\label{decays}
\end{figure}

The strong decays of hadrons has been, and remains, a mystery of soft QCD. The naive
perturbative assumption that the decay proceeds via one gluon dissociation 
(Fig. \ref{decays}b) is proven incorrect by direct comparison with 
experiment\cite{Geiger:1994kr}. The only reasonably successful phenomenology
is provided by the `$^3P_0$' model\cite{3p0}, where quark pairs are assumed to appear
with vacuum quantum numbers over all space. This is clearly an unacceptable situation,
especially given the ubiquity of hadronic decays and the fact that they provide a 
window into the dynamics of glue at low energy.

We now examine the predictions of the new quark model presented here. To
lowest order in $\Lambda_{QCD}/m_g$ 
 and to all orders in the coupling, the
 only diagrams which contribute to meson decay (here all mesons are
 assumed 
 to have Fock expansions which are dominated by the quasiquark-quasiantiquark components)
are shown in Fig. \ref{decays}.  
 The left figure is contained within $H_0$ and is therefore
the leading diagram. The central and right figures contribute at ${\cal O}(\Lambda_{QCD}/m_g)$ and are generated by $H_{int}$.

Diagrams (a) and (b) with perturbative gluons or model potentials in the intermediate
 states have been previously examined as possible sources
of hadronic decays in Ref.~\cite{Ackleh:1996yt}. The authors noted that diagram (a) is strongly
suppressed with respect to (b) due to momentum routing through the pair production
vertex (this diagram is zero in the nonrelativistic limit when a 
delta function
potential is in place. It is strongly suppressed with a $1/q^4$ potential). The other
class of diagrams considered in Ref.\cite{Ackleh:1996yt} was that generated by a
phenomenological scalar interaction given in terms of scalar confinement (cf. Eq. (\ref{naive})).  This is, of course, an {\it ad hoc} microscopic realization of the $^3P_0$ model. What was
found was that this diagram (like diagram (a) but with scalar as opposed to a vector
vertices) was much larger than diagram (b).

These conclusions imply that the $^3P_0$ model would emerge in a natural way from 
our methodology if diagram (c) produced light quark pairs with scalar quantum
numbers. Diagram (c) is generated by the product of $K^{(1)}$ and $H_{qg}$ terms in 
$H_{int}$ (see Eqs. (\ref{hqg}) and (\ref{kexp})) and is roughly given by 
$\A\cdot\bnabla \psi^\dagger \alpha \cdot \A \psi$.
Once the vector potentials are contracted (or better yet, the sum over intermediate
hybrid states is made), the resulting operator is of the form $\psi^\dagger 
[\bsigma \delta_T \bdel] \psi$,
very nearly equal to the long-assumed $^3P_0$ vertex. Thus we 
have obtained a viable microscopic description of hadronic decays.  The implications
of these observations will be explored in a future publication.

\subsection{Light Quarks and the Constituent Quark Model}

%

The utility of the CQM for heavy quarkonium is not in doubt. However,
its apparently successful extension to light quark states is
unexpected and surprising. We seek to understand this observation in this subsection.

\begin{figure}[hbp]
\hbox to \hsize{\hss\psfig{figure=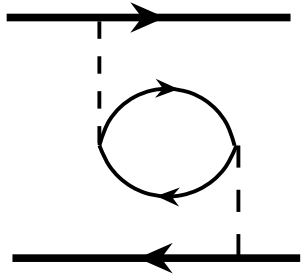,width=2.5in,angle=0}\hss}
\vspace{0.5cm}
\caption{Light Quark Loop Correction to the Heavy Quark-Antiquark Interaction.}
\label{Fig14}
\end{figure}

The major feature of light quark physics is spontaneous chiral symmetry breaking. 
One may regard this as occurring due to the appearance of a quark-antiquark vacuum
condensate. The interactions which generate the condensate are typically associated
with an effective instanton interaction\cite{ssreview} or the confinement 
potential\cite{Finger:1982gm,Adler:1984ri,LeYaouanc:1985dr}. (In our approach
the driving kernel would be $K^{(0)}$). Regardless of the particular mechanism which
causes attraction in the scalar channel, a massive constituent quark is the 
necessary outcome. Indeed,
while bare quarks may become very light
or massless, the relevant quasiparticles saturate at roughly 200 MeV as the
bare quark mass is reduced\cite{Szczepaniak:1997gb,Szczepaniak:2000bi}.
This, at least partly, explains the
apparent success of the nonrelativistic portion of the CQM.
The agreement is also enhanced by the empirical accident that the expectation value
of $\sqrt{p^2 +m^2}$ is very close to $\p^2/2m_{\rm CQM}$ in typical hadronic states.
More important than this; however, is the nature of the central potential
itself when the bare quarks are light. As discussed above, effects due to one
gluon exchange are suppressed by powers of $\Lambda_{QCD}/m_g$. 
Thus the main effect due to light
quarks is the presence of intermediate quark loops in the instantaneous interaction
(Fig. \ref{Fig14}).
These diagrams cause string breaking which is an important feature of QCD.
However, as Isgur has argued\cite{Isgur:1999cd},
the main consequence of this is simply to renormalize the string tension.
Thus light quark loops have little effect on the phenomenology arising
from $H_0$. The conclusion is that structure of the new CQM which
we have laid out is essentially unchanged for light quarks. Furthermore,
even the simple nonrelativistic approximation may retain some validity
for massless bare quarks.

An explicit demonstration of how the CQM may emerge was given in 
Ref.~\cite{Szczepaniak:2000bi}. This paper assumed a simple contact interaction
in place of the full Coulomb kernel.
Standard many-body techniques were used to obtain
chiral symmetry breaking and constituent (quasiparticle) quarks. It was then 
demonstrated that the vector meson -- pseudoscalar meson mass splitting follows a form
essentially identical to that of the CQM hyperfine splitting when considered as  a function
of the constituent mass.  Nevertheless, the mass splitting was clearly driven by
chiral symmetry breaking when considered as a function of the current quark mass, thereby
demonstrating that the pion may be viewed as both a pseudoGoldstone boson and as a 
quark-antiquark bound state. The new quark model presented here provides an explicit 
microscopic realization of
the contact model employed in Ref.~\cite{Szczepaniak:2000bi} and it will be of interest 
to verify the findings of that work.


\section{Conclusions}

In the paper we propose a new way to organize QCD which is appropriate for low
energy hadronic physics. The starting point is chosen to be the QCD
Hamiltonian in Coulomb
gauge because this gauge is most directly applicable to bound state physics -- the
degrees of freedom are physical and an instantaneous potential exists\footnote{We note that 
it is also useful for QCD at finite temperature because a special frame is automatically selected and 
because counting
degrees of freedom is an important aspect of thermodynamics.}. The instantaneous Coulomb
potential may be incorporated into $H_0$ (as is done in atomic physics) and a viable
bound state perturbation theory may be constructed. This simple step already obviates one
of the severe problems of perturbative QCD in describing hadronic properties, namely that of
ill-defined asymptotic states.

While the division of the QCD Hamiltonian is a simple task, it is essentially
meaningless 
because the degrees of freedom represented
in $H_0$ are partonic. Thus building bound states would be a frustrating exploration of
the depths of Fock space rather than the preliminary step for bound state perturbation
theory we desire it to be.  The experience provided by the constituent quark model
points the way out of the impasse: appropriate (constituent) 
degrees of freedom must be employed. The problem in the past (cf. constituent quark models,
bag models, flux tube models, etc) has been in finding a way to introduce effective degrees of 
freedom in such a way that the connection to QCD is not destroyed. Herein we present one
way to do this which is based on experience gleaned from many-body
physics often used in phenomenological models {\it e.g.} 
the Nambu--Jona-Lasinio model\cite{NJL}. Specifically, a canonical
 transformation to a quasiparticle
basis which is defined with respect to a nontrivial variational vacuum is made. The theory
remains QCD but is given in terms of a useful and tractable basis. Although the
vacuum state is necessarily an ansatz, this does not vitiate the construction -- in principle
{\it any} basis may be used, we merely seek an efficient one, and the vacuum itself may
be systematically improved with standard techniques.

One finds a welcome complication when these ideas are applied to nonabelian gauge theory: the
interaction  which is needed to define the vacuum ansatz and the quasiparticle spectrum
(via the gap equation) itself depends on the vacuum. Thus the fundamental quasiparticle interaction
and the quasiparticles themselves are inextricably interdependent. Solving the gap equations
requires the evaluation of the specific functional dependence of the quasiparticle interaction 
on the vector potential. We have chosen to do this within the rainbow ladder approximation. 
There are several important points to make at this stage: (1) the rainbow ladder approximation
may be improved at will, (2) the approximation is accurate in the large $N_c$ limit,
(3) the approximation is accurate in the infrared limit, (4) the approximation is 
justified {\it a posteriori}. Lastly, although the approximation cannot 
yield nonperturbative results, true nonperturbative physics may be generated
when the resummed kernel is incorporated in the nonlinear coupled gap equations. 
Doing so reveals a pleasant surprise: the emergence of the confinement phenomenon.

While it is gratifying that color confinement is produced by our
approach, this result would
be useless if it did not match phenomenology. The fact that the effective
quasiparticle potential matches the lattice static quark potential very well
points to the general utility of our method.  Thus Eqs.~(\ref{h0}) and (\ref{hint})
represent much more than a simple reordering of the
QCD Hamiltonian. By building $H_0$ as an effective Hamiltonian describing the
interactions of quasiparticles on a nontrivial vacuum we are able to establish
contact with the constituent quark model and derive confinement.  That
both of these emerge in our formalism bodes well for the future success of $H_0$
as a robust starting point for detailed hadronic computations.

An important test of any new method in QCD is its ability to 
provide insight into a variety of phenomena. We have tried to demonstrate
the robustness of our method in this regard. A vital aspect of this
robustness is the emergence of $\Lambda_{QCD}/m_g$ as an expansion parameter.
This provides the justification for gluonic Fock space truncation, for the
validity of the leading static Coulomb kernel $K^{(0)}$, and for the
applicability of the static kernel to light quarks. Indeed, the method
strongly hints as to why the constituent quark model works for light quarks.
To summarize, quarks never become truly light (but saturate at constituent masses),
the static kernel is not strongly affected by the presence of light quarks, and
parameter freedom in the definition of the quark model allows for an accurate
reproduction of the relativistic quark kinetic energy and the chirally-driven 
meson hyperfine splitting.

The ideas we have presented have had a long period of development starting with
Gribov's speculation that confinement may arise naturally when resolving the gauge
copy problem.
In the early 1980's Finger and Mandula\cite{Finger:1982gm}, 
Adler and Davis\cite{Adler:1984ri}, 
and Le~Yaouanc {\it et al.}\cite{LeYaouanc:1985dr} all considered the generation of constituent 
quark masses and
spontaneous chiral symmetry breaking with simple (often of the form given in Eq.~\ref{naive}) 
models of QCD. The issue of renormalization was taken up by these papers and in 
Refs.~\cite{Szczepaniak:1997gb,Szczepaniak:2000uf,Robertson:1999va}.  

The work which is closest
to ours is that of Zwanziger\cite{Zwanziger:1998ez} and Swift\cite{Swift:1988za}. 
As discussed in Sec. IV.B, Zwanziger has shown that the imposition of 
the horizon condition
implies that the Faddeev-Popov propagator is enhanced in the
infrared. As we have stressed, an enhancement of the FP propagator is 
sufficient to cause confinement. In 
Ref.~\cite{Zwanziger:1998ez}  Zwanziger  
has shown that adding the horizon term to the Hamiltonian 
produces an effective gluon mass which in turn induces the desired
enhancement of the VEV of the Faddeev-Popov operator. 
Zwanziger then makes several simplifying assumptions to arrive at an estimate for the
Coulomb kernel.  Chief among these are an assumed form for
the gluon dispersion relation, a simplified version of the Faddeev-Popov propagator
integral equation, and the approximation $K \sim d^2(k)/k^2$.
The end results are similar to ours; Zwanziger obtains 
$d \sim k^{-4/3}$ (we get $k^{-1/2}$) and $V \sim r^{5/3}$.
Our analytical approximation gives $V \sim r^{3/4}$ while the numerical solution is very
nearly linear. 

The work of Swift\cite{Swift:1988za} is very similar to ours in philosophy. 
In fact our self-consistent equations for the leading rainbow-ladder
gap equations, which were derived in the Hamiltonian formalism,  agree with 
those of Ref.~\cite{Swift:1988za}, which were derived in the Greens function formalism.
However, a difference occurs in the renormalization
of the mass gap equation: we find that only one subtraction is necessary to render the
equation finite. Thus no counterterm proportional to $\A \nabla^2 \A$ is required.
This is due to the logarithmic suppression of the potential at large momenta.

The main difference between the current paper and Ref.~\cite{Swift:1988za} is in
the analysis of the gap equations. We have obtained  very good 
analytic and 
full numerical solutions to the coupled gap equations. 
This was not attempted in Ref.~\cite{Swift:1988za}; however, the author did examine
the small momentum behaviour of the
Faddeev-Popov 
propagator and the Coulomb kernel
assuming a particular form for the  gap function. 
His FP and gap functions agree with our
analytical estimates, however his solution
for the potential has an unexpected imaginary portion.
We believe this is due to an approximation which generated a confinement potential
which was more singular than $1/k^4$.

A preliminary exploration of the work presented here was undertaken in Ref.~\cite{Szczepaniak:2000uf}.
This reference neglected the FP determinant  and higher order contributions to the
gap equation. Furthermore, the full Coulomb kernel was drastically simplified by
taking it to be the summation of the one loop expression for $K$. Despite these
simplifications and assumptions the resulting potential was similar to that obtained
here. This is perhaps an indication of the power of the coupled quasiparticle/vacuum 
approach.

We regard the present work as a promising start to the construction of a
new quark model of the strong interactions in particular with regard
to the treatment of gluonic degrees of freedom. 
Future projects include the
evaluation of all $d^4$ correction terms to the gap equation which are needed
to test the Gribov-Zwanziger gluonic quasiparticle spectral
function. We also intend to evaluate 
a broad swath of the meson,  baryon, glueball, and hybrid  spectra,
and to compute the heavy hybrid
adiabatic energy surfaces. The latter two test the utility of the
gluonic quasiparticles as effective degrees of freedom and will probe the
structure of $K^{(0)}$ and $K^{(1)}$. General considerations (and
explicit lattice evidence) lead one to expect that glue behaves as a collective
stringlike degree of freedom at large distances. We expect the gluonic 
quasiparticles to provide a useful description of glue at small (less
than 1 fermi) and intermediate (1-2 fermi) scales. It will be interesting
to see if the formalism presented here allows for effective
stringlike behavior at large distances. Finally, $\eta-\eta'$
mixing is a longstanding issue in soft QCD  related to the anomaly,
vacuum structure, the gluonic content of mesons, and instantons.  Examining this problem
should prove informative for the further development of the quark sector of
our theory.

\begin{acknowledgements}
We would like to thank R. Alkofer,  S. Brodsky, A. Duncan, N. Isgur,
E. Shuryak, H. Thacker,  A. Thomas, and D. Zwanziger
for discussions. 

This work was supported by DOE under contracts
DE-FG02-00ER41135,  DE-AC05-84ER40150 (ES), and DE-FG02-87ER40365 (AS).
\end{acknowledgements}

\end{document}